\newif\ifpreprint
\preprinttrue
\ifpreprint
\documentclass[11pt]{article}
\else
\documentclass[sn-mathphys-num]{sn-jnl}
\fi

\usepackage{graphicx}%
\usepackage{multirow}%
\usepackage{amsmath,amssymb,amsfonts}%
\usepackage{amsthm}%
\usepackage{mathrsfs}%
\usepackage[title]{appendix}%
\usepackage{textcomp}%
\usepackage{manyfoot}%
\usepackage{booktabs}%
\usepackage{algorithm}%
\usepackage{algorithmicx}%
\usepackage{algpseudocode}%
\usepackage{listings}%
\usepackage{mathabx}

\usepackage[table,xcdraw]{xcolor}
\usepackage{booktabs}
\usepackage{multirow}
\usepackage{array}
\usepackage{geometry}

\definecolor{rank1}{RGB}{0,102,0}     
\definecolor{rank2}{RGB}{34,139,34}   
\definecolor{rank3}{RGB}{255,165,0}   
\definecolor{rank4}{RGB}{255,99,71}   
\definecolor{rank5}{RGB}{220,20,60}   
\definecolor{rank6}{RGB}{139,0,0}     

\newcommand{\rankcolor}[1]{%
  \ifnum#1=1\textcolor{rank1}{#1}\else%
  \ifnum#1=2\textcolor{rank2}{#1}\else%
  \ifnum#1=3\textcolor{rank3}{#1}\else%
  \ifnum#1=4\textcolor{rank4}{#1}\else%
  \ifnum#1=5\textcolor{rank5}{#1}\else%
  \ifnum#1=6\textcolor{rank6}{#1}\else%
  #1\fi\fi\fi\fi\fi\fi%
}

\usepackage         {algpseudocode}
\usepackage         {algorithm}
\usepackage         {amsfonts}
\usepackage         {amsmath}
\usepackage         {amssymb}
\usepackage         {amsthm}
\usepackage         {bbm}
\usepackage[british]{babel}
\usepackage         {booktabs}
\usepackage         {comment}
\usepackage         {dsfont}
\usepackage         {mathtools}
\usepackage         {microtype}
\usepackage         {nicefrac}
\usepackage         {paralist}
\usepackage         {subcaption}
\usepackage         {siunitx}
\usepackage         {wrapfig}
\usepackage         {thm-restate}

\ifpreprint
\usepackage[square,numbers]{natbib}
\usepackage{geometry}
\else
\usepackage{natbib}
\fi

\usepackage{enumitem}
\usepackage{rotating}
\newcommand{\reals}{\mathds{R}}
\newcommand{\ones} {\mathds{1}}
\usepackage{hyperref}
\usepackage[capitalise]{cleveref}

\DeclareMathOperator{\clique}{\mathrm{Cl}} 
\DeclareMathOperator{\PD}{\mathcal{D}} 
\DeclareMathOperator{\TL}{\mathrm{TL}} 
\DeclareMathOperator{\B}{\mathrm{b}}

\begin{document}


\ifpreprint

\title{Detecting Spatial Dependence in Transcriptomics Data  
using Vectorised Persistence Diagrams}

\date{}

\author{
  Katharina Limbeck\\
  \small{Helmholtz Munich, Technical University of Munich}\\
  \texttt{katharina.limbeck@helmholtz-munich.de}
  \and
  Bastian Rieck\\
  \small{Helmholtz Munich, Technical University of Munich, University of Fribourg}\\
  \texttt{bastian.grossenbacher@unifr.ch}
}

\makeatletter
\let\@mkboth\@gobbletwo
\let\@oddhead\@empty
\let\@evenhead\@empty
\makeatother

\fi

\maketitle

\ifpreprint
\maketitle

\begin{abstract}
Evaluating spatial patterns in data is an integral task across various domains, including geostatistics, astronomy, and spatial tissue biology.
The analysis of \textit{spatial transcriptomics data} in particular relies on methods for detecting spatially dependent features that exhibit significant spatial patterns for both explanatory analysis and feature selection.
Given the complex and high-dimensional nature of these data, there is a need for robust, stable, and reliable descriptors of spatial dependence.
We leverage the stability and multi-scale properties of \textit{persistent homology} to address this task.
To this end, we introduce a novel framework using functional topological summaries, such as Betti curves and persistence landscapes, for identifying and describing non-random patterns in spatial data.
In particular, we propose a non-parametric one-sample permutation test for spatial dependence and 
investigate its utility across both simulated and real spatial omics data. 
Our vectorised approach outperforms 
baseline methods at accurately detecting spatial dependence. 
Further, we find that our method is more robust to outliers than alternative tests using Moran’s I. 
\end{abstract}
\fi

\section{Introduction}\label{sec1}

Omics data, that is measurements of molecules within an organism or a cell, provide crucial insights into biological processes and cellular function \citep{vandereyken2023methods}. 
Specifically, transcriptomics plays a vital role in studying RNA transcripts, i.e. the expression of genes via the transcriptome. As a fundamental building block of molecular biology, transcriptomic measurements provide snapshots of cellular function and the mechanisms behind health and disease \citep{palla_spatial_2022}. 
Recent advancements in spatial transcriptomics further enable the large-scale analysis of cells’ gene expression and their spatial locations across samples or within a tissue. 
Leveraging these data, the detection of spatially variable genes (SVGs) is an important analysis step. Spatially dependent transcripts are used both for explanatory analysis and for further downstream tasks, such as spatial domain identification and gene enrichment analysis  \citep{palla_spatial_2022, heumos2023best}. Thus, SVG detection is crucial for uncovering  the spatial components of tissue biology. 
However, accurately identifying the spatial patterns in 
 omics data on both local and global scales remains challenging, particularly due to the high-dimensional, sparse, and noisy nature of these data \citep{amezquita2020shape}. Challenges include the difficulty of modelling  
technical and systematic errors, the \mbox{ongoing} debate on even basic preprocessing steps, and an ever-growing understanding of the fundamental processes in cellular biology \citep{heumos2023best}. These hurdles necessitate the  development of robust and expressive analysis methods \citep{rabadan2019topological}. 
In particular, SVG detection methods should be robust to technical noise and sparsity, correctly identify spatially dependent genes, be independent of the sum of  expression values, and show low false-positive error rates \citep{chen2024evaluating, li2023benchmarking}. In practice, reliable SVG detection methods then help identify the spatial patterns of disease. For example, the analysis of spatially dependent transcripts enables the detection of spatial clusters of cancer cells or necrotising tissue within an organ. Further, SVG detection helps identify the genes driving these spatial changes \citep{palla_spatial_2022}.

Topological data analysis (TDA) has given scholars a new toolbox for quantifying complex structures in biological networks at multiple scales, motivated by the rigorous mathematical study of shapes \citep{feng_spatial_2020}. 
Indeed, when adjusting for noise, molecular data in particular is equipped with meaningful coarse geometry 
\citep{rabadan2019topological}. 
TDA is thus poised 
to quantify spatial patterns in molecular data by capturing both local and global characteristics of the shape of data. 
Computing multi-scale topological features on spatial graphs gives stable and flexible summaries of the distribution of spatial features \citep{byers_detecting_2023}, summarising both local and global patterns in gene expression.  
In particular, our work demonstrates how topological descriptors can be used to distinguish spatial signals in transcriptomics data and identify genes whose expression varies across space.
Our \textbf{contributions} are as follows:
\begin{itemize}
    \item We propose a one-sample randomised permutation test for spatial dependence using functional topological summaries. 
    \item We investigate the performance of this approach on both simulated and real data in comparison to alternative methods, Moran's I, \texttt{sepal} and \texttt{SpatialDE}.
\end{itemize}

\section{Related Work}

\subsection{ 
Spatially Variable Gene Detection} 

The study of omics data underpins our understanding of fundamental biological processes \citep{vandereyken2023methods}. The term `omics' denotes molecular measurements in general, including features extracted from the genome, transcriptome, or proteome. 
As a key mechanism in biology, the set of DNA encoded in the genome is read out or transcribed by RNA molecules, which make up the transcriptome. Some of these transcripts later code for proteins, which are of great practical importance as they perform the actual biological functions in cells. Omics measurements then give key insights into key aspects such as disease pathology, cellular interactions, and disease development \citep{palla_spatial_2022, reel2021using}. 
In the last decade, rapid advancements in  sequencing technologies have enabled the rapid growth in the collection and analysis of spatial omics data \citep{moses2022museum}. 
Spatial data are particularly interesting as they provide information on both molecular measurements as well as spatial tissue organisation. 
Observations from spatial omics datasets either correspond to individual cells, enabling the advent of single cell omics, tissue regions spanning multiple cells, or even sub-cellular measurements. Associated spatial locations are given by unstructured, i.e. continuous coordinates on a tissue sample, 
or spots on a pre-defined grid \citep{palla_spatial_2022}. 
The focus of our investigation in particular are transcriptomic measurements that  summarise the number of times a gene is measured by the number of RNA fragments detected at each capture location.

As a key processing step, transcriptomics data allows practitioners to detect genes whose expression varies notable between dissociated cells as the most highly variable cells of interest \citep{heumos2023best}. Spatial transcriptomic further allows us to determine genes whose expression varies in space. Such spatially variable genes then give key information on biological processes driving spatial tissue architecture, which motivates the development of a range of computational methods for detecting spatially variable genes~\citep[SVGs]{palla2022squidpy}. 

Generally, SVG detection methods fall into numerous categories. Moran's I is the most established statistical measure for spatial autocorrelation across spatial statistics \citep{moran1950notes}. As well as its cousin Geary's C \citep{geary1954contiguity}, Moran's I is frequently used as a versatile descriptor of spatial patterns in gene expression values \citep{heumos2023best, li2023benchmarking}. 
Despite its simplicity as a mean summary based on weighted variance and covariance estimates, and its well-established theoretical benefits, 
Moran's I is 
known to be sensitive to outliers and noise \citep{tiefelsdorf1997note, bucher2020estimation}, a disadvantage given the noisy  nature of spatial omics data \citep{chen2024evaluating}. 
Further, based on the rich tradition of spatial statistics, there also exists a variety of statistical methods that fit spatial regression models to assess the importance of spatial covariance on gene expression. The most  popular of these approaches include \texttt{SpatialDE} and \texttt{SpatialDE2}, which fit Gaussian regression processes, and \texttt{SPARK} or \texttt{SPARK-X}, fitting mixed and non-parametric models, respectively \citep{chen_evaluating_2022}. However, these models rely on fixed assumptions on the nature of the distribution of gene expression values, which remains a contested question in practice \citep{zhu2023srtsim}. 
Lastly, there exist non-parametric methods, such as 
\texttt{trendsceek}, which models gene expression as 
a marked point process and uses permutation testing to make inferences \citep{edsgard2018identification}. This approach benefits from allowing statistical reasoning under less stringent assumptions but is associated with increased computational costs. 
Further, 
model-free frameworks, such as \texttt{sepal} \citep{andersson2021sepal}, which measures diffusion times, or \texttt{GLISS}, which models graph Laplacians, have also been proposed \citep{adhikari2024recent}. 
However, despite the appeal of such geometric and model-free approaches, multi-scale descriptors, such as persistent homology, have not yet been employed in this context. 
Motivated by the use of geometric descriptors of spatial graphs, we thus aim to further develop persistent homology as a framework for spatial significance testing and focus on applications to spatial omics data.

\subsection{Persistent Homology}

Originating from algebraic topology, persistent homology (PH) enables practitioners to measure topological features from data \citep{edelsbrunner2008persistent}, thus describing the \textit{shape of data} \citep{amezquita2020shape}. 
This is of particular importance for molecular data, which, after accounting for noise, exhibits meaningful coarse geometry \citep{rabadan2019topological}. PH then provides one such a tool for encoding this geometry in a robust manner.

From a signal-processing perspective, persistent homology has demonstrated its utility for detecting peaks in noisy signals, guaranteeing stability properties and robustness to small perturbations \citep{huber2021persistent, hickok2022analysis, feng_spatial_2020}. The multi-scale nature of filtrations allows practitioners to summarise not only local but also global patterns in spatial data. 
Persistence-based methods can thus be applied to detect and encode both local and global extrema alike \citep{hickok2022analysis}.
Indeed, the use of persistent homology as spatial descriptor has been extended to investigating time-varying data and spatial clustering \citep{hickok2022analysis, pereira2015persistent, Rieck16a}. 
Thus, persistence has shown its utility for describing spatial patterns, such as local maxima and local minima, in various contexts. For these reasons, persistence shows especially  promising potential for SVG detection.

From a statistical perspective, 
the increasingly widespread use of PH for data analysis and budding applications to machine learning have also been driven by advancements in the statistical theory, underpinning the theoretical foundation of PH. In particular, summaries of PH allow for the definition of parametric and non-parametric significance tests and uncertainty estimation. Amongst these approaches, non-parametric techniques, specifically permutation testing, are by far the most utilised and assumption-free methods used for statistical inferences \citep{bubenik2015statistical,robinson_hypothesis_2017}. 
Arguably, one-sample testing remains somewhat underexplored \citep{byers_detecting_2023}, with most common statistical approaches focusing on two-sample comparisons.

Nevertheless, recent work by \citet{byers_detecting_2023} demonstrates the use of persistent homology 
for the identification of spatially dependent features in geographical data. In particular, they propose a one-sample permutation testing approach using persistence summary statistics  
\citep{byers_detecting_2023}. 
Their findings suggest that 
persistent homology performs well at testing for spatial dependence in the presence of outliers or on sparsely connected graphs resembling ladder graphs \citep{byers_detecting_2023}. 
Further, they conclude that there is potential for future work to investigate the use of alternative functional summaries of persistent homology. 
Addressing this outlook and widening the application to spatial omics data, we now investigate the use of persistence curves for SVG detection.

\section{Background on Persistent Homology}

\label{sec:ph}
\subsection{Spatial Data as Spatial Graphs}
\label{sec:1data}
\begin{figure}[tbp]
    \centering
    \includegraphics[clip, trim=0cm 8cm 0cm 0cm, width=1\textwidth, page=2]{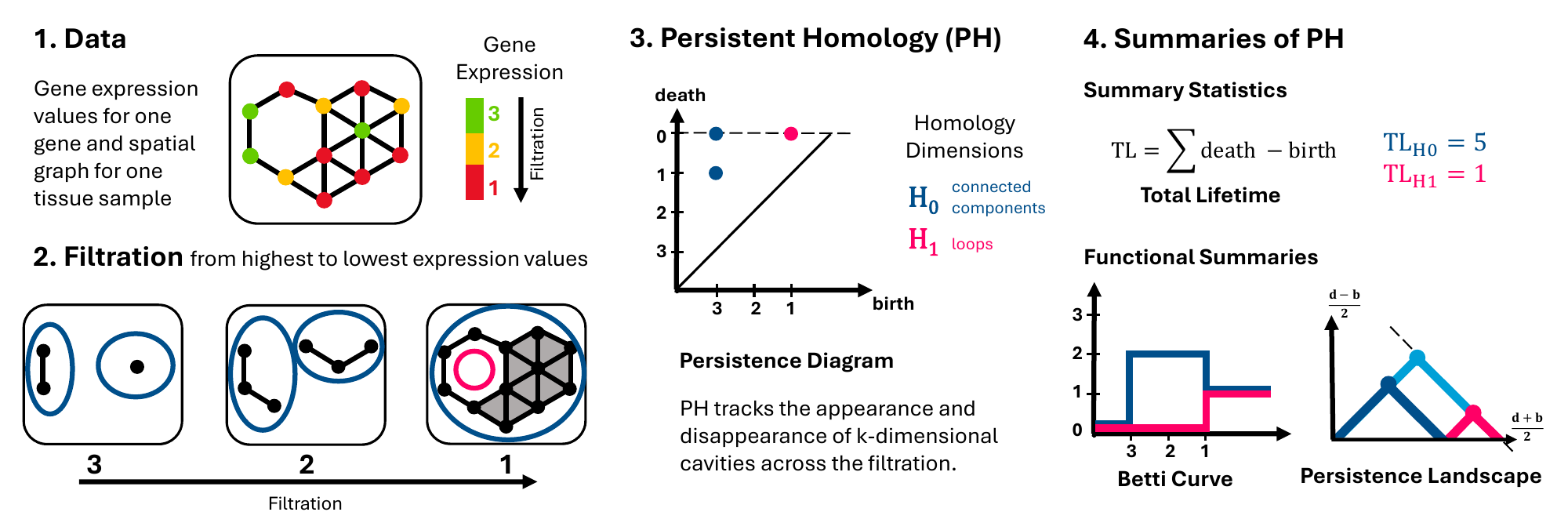}
    \caption{Overview of the proposed analysis pipeline (1)
    starting from a spatial omics dataset that is converted to a graph, (2) defining a superlevel-set filtration based on the expression values of one gene, (3) computing persistent homology by tracking the evolution of topological features via persistence diagrams, and (4) summarising persistence based on summary statistics or functional summaries.}\label{fig:over}
\end{figure}

Throughout this paper, we study 
a set of observations \(X = \{x_1, x_2, ..., x_n\}\), which correspond to cells or spatial regions on a tissue sample. Each observation is further associated with a spatial location \(l(x_i) \in \reals^2 \text{ for } x_i \in X\). We are interested in assessing the spatial distribution of one gene or feature \(f(x_i) = f_i \in \reals \text{ for } x_i \in X\). This feature corresponds to measurements of one specific mRNA molecule of interest across a tissue slice. That is, the value \(f_i\) measures the level of gene expression of one specific transcript at a spatial observation \(x_i \in X\). 
We then want to examine the spatial distribution of this gene, which is modelled by constructing a spatial (neighbourhood) graph \(G = (X, E)\), so that each vertex corresponds to an observation in \(X\) and the edge set is constructed from the spatial locations. For example, an  \(\epsilon\)-neighbourhood graph can be constructed as \begin{equation}E=\{(x_i, x_j) \in X \times X \mid  d(l(x_i), l(x_j)) \leq \epsilon\},
\end{equation} 
where \(d: \reals^2 \rightarrow \reals\) is a spatial distance metric between observations and \(\epsilon \in \reals\) is some neighbourhood radius. 
This approach is 
appealing because adjacency relationships play a central role in spatial data analysis. By choosing a suitable neighbourhood criterion, 
we can flexibly construct a graph that best describes the nature of the data. See \cref{sec:omics_data} for a more detailed description on our method. As an example illustrating our methodology, \cref{fig:over}.1. shows a spatial neighbourhood graph, where the colour of each vertex indicates the expression values of one exemplary gene measured at that spatial location. 

\subsection{Persistent Homology and Vertex-based Filtrations}
Persistent homology can be computed from a spatial graph \(G=(X,E)\) 
by considering the clique-complex \(S=\clique(G)\) defined as the simplicial complex made up of cliques of vertices. Here, a clique complex is a generalisation of a graph that consists of 0-simplices corresponding to vertices, 1-simplices corresponding to edges, 2-simplices corresponding to 3-cliques~(i.e.\ triangles), and so on. More formally, a \(k\)-dimensional simplex in \(S\) corresponds to a set of \(k+1\) vertices that form a complete subgraph. Specifically, we have
\begin{equation}
    \clique(G) = \{ \sigma \subseteq X \mid G[\sigma] \text{ is a complete graph} \},
\end{equation}
where \(G[\sigma]\) is the subgraph of the graph $G$, induced by the vertex set \(\sigma\).
The clique complex defines a simplicial complex, i.e. a set of simplices that is closed under taking intersections and face decomposition, as any subset of a clique is also a clique itself. 
In practice, we can think of \(S\) as a generalised triangulation of a spatial graph. 

As a multi-scale summary of geometry, persistent homology then studies a \emph{filtration} on this complex. That is, we construct a sequence of topological spaces, and track the evolution of topological features given by homology groups. 
The added algebraic structure allows for the computation of boundaries and holes from the generalised graph. We can thus track the emergence and disappearance of connected components, that is zero-dimensional homology groups, \(H_0(S)\), and more generally, \(k\)-dimensional homology groups, \(H_k(S)\), representing \(k\)-dimensional holes \citep{huber2021persistent}. This process, as well as the further details explained below, are illustrated via an example in \cref{fig:over}.2. which plots the evolution of connected components (in blue) and one-dimensional loops (in pink) across a filtration on a spatial graph. 

While there exists a multitude of methods for defining filtrations on graphs \citep{aktas2019persistence}, we are interested in one specific construction throughout this work. In particular, we track the evolution of superlevel-sets defined from the spatial vertex feature \(f\colon X \rightarrow \reals\), as previously defined in \cref{sec:1data}. This filtration allows use to investigate the spatial distribution of one gene as we vary from the highest to the lowest expression value. 
That is, we filter through the simplicial complex as follows
\begin{equation}
    S_{\delta} = \clique(G_{\delta})= \{s \in  Cl(G) \mid f(x_i) \geq \delta \text{ for all } x_i \in s\}
    \label{eq:filt}
\end{equation}
so that the filtration parameter \(\delta \in [\max(f_i), \min(f_i)]\) decreases from the maximum to the minimum feature value. Here, \(s\) is a simplex in the clique complex \(S_{\delta}\) of the subgraph \(G_\delta \in G\) defined as \(G_\delta = G[\{x_i \in X \mid f(x_i) \geq \delta\}]\). 
%
The collection of simplicial complexes \(S_{\max(f_i)} \subset ... \subset S_{\min(f_i)} = Cl(G)\) then gives a 
filtration of nested simplicial complexes. 
Intuitively, this vertex-based filtration can be constructed by 
filtering through 
the feature values starting with an empty simplicial set for \(\delta > \max(f_i)\). 
Then, as \(\delta\) decreases, a vertex  \(x_i \in X\), i.e a zero-dimensional simplex, is included whenever its associated vertex feature value \(f(x_i)\) has been reached. Further, an edge, i.e. a one-dimensional simplex, is added to the filtered simplicial complex \(S_{\delta}\) whenever both its vertices have been included \citep{aktas2019persistence, rieck2017clique}, with the process generalising to higher dimensions. \cref{fig:over}.2. illustrates an example of a superlevel-set filtration constructed in this manner. 
Note that to define a sublevel-set filtration, this construction can be modified, so that \(\delta\) increases from \(\min(f_i)\) to \(\max(f_i)\). Further, for data collected on a rectangular grid, cubical persistence, which filters through cubical complexes made up of cubical cells, offers a natural alternative to the triangle-based simplicial complex construction. See \citet{wagner2011efficient} for more details on how to generalise to this setting.

The output from computing these filtrations results in a set of persistence diagrams, \(\{\PD_0, \PD_1\}\), one for each homology dimension \(k\), where each \(\PD_{k} = \{ (p_1, q_1), (p_2, q_2), ..., (p_m, q_m)\}\) is a multi-set of persistence pairs in $\reals^2$ encoding the birth \(p_i\) and death \(q_i\), i.e.\ the emergence and merging scales of topological features in the filtration. 
In particular, peaks or local maxima in spatial feature values are represented by points in the 0-th dimensional persistence diagram above the diagonal \citep{huber2021persistent}. This is illustrated in an example of a persistence diagram in \cref{fig:over}.3., where the two persistence pairs plotted in blue correspond to the connected components that emerge during the filtration from the highest to the lowest gene expression values. The persistence diagram thus encodes spatial clustering behaviours in gene expression values which we want to leverage for detecting spatial dependence. 
For a deeper exploration of the mathematical framework and theoretical underpinnings of persistent homology, we refer the interested reader to  \citet{edelsbrunner2008persistent}. 
We, however, move on to consider the use of persistent homology for statistical analyses.


\subsection{Summarising Persistence Diagrams} 

Persistence diagrams consist of multi-sets of birth and death pairs and are thus challenging to study from a statistical perspective. In fact, there exists no natural way to directly understand them in a Hilbert space that allows for the application of standard statistical tools. Moreover, inherent distances between diagrams such as the Bottleneck or Wasserstein distances are expensive to compute.
Thus, to gainfully use PH in a statistical setting, we can use  summary statistics, as for example explored by \citet{byers_detecting_2023} in the context of spatial dependency detection. Moreover, to retain the multiscale properties of PH, functional summaries can be employed to understand persistence diagrams in a Hilbert space setting. \cref{fig:over}.4. illustrates the examples of three of such persistence summaries, which we will now explain in more detail.

\subsubsection{Summary Statistics}
While various one-number descriptors such as persistence entropy or lifetime-derived statistics are available, we choose to include total lifetime (TL) into our study based on the findings of \citet{byers_detecting_2023}. 
In particular, the total lifetime is computed as: 
\begin{equation}
    \TL = \sum_{i=1}^m (q_i - p_i).
\end{equation} That is, total lifetime is the sum of the lifetimes of persistence pairs, defined as the difference between the death \(q_i\) and birth \(p_i\) scales of topological features \((p_i, q_i) \in \PD\). 

\subsubsection{Functional Summaries}
Our analysis will focus on using functional summaries that map persistence diagrams to (piecewise) continuous functions from \(\reals \to V\) where \(V\) is a suitable vector space \citep{ali2023survey}. Note that in related literature these types of functional summaries are often called persistence curves. In particular, we compare Betti curves 
and persistence landscapes, which can be conveniently compared via suitable norms between these functions. We hypothesise that these functional summaries retain more complete information from the whole persistence diagram than single number summaries (as e.g. studied by \citet{byers_detecting_2023}) while allowing for the application of statistical methods in a theoretically well-founded manner \citep{bubenik2015statistical, rieck2020topological}. 

\paragraph{Betti curves}
Betti curves are defined as 
$\B\colon \reals \to \mathbb{N}$ where 
\begin{equation}
    \B(\delta)=|\{ (p_i, q_i) \in \PD \mid \delta \in [p_i, q_i]\}|.
\end{equation} That is, Betti curves sum up the number of topological features from the persistence diagram that persist at each scale of the filtration. Specifically, we filter from the highest vertex feature value, \(\max(f_i)\) to \(\min(f_i)\). 
Given one Betti curve, \(b\), we can then take its \(L^p\) norm as a natural descriptor:
\begin{equation}
    L^p(\B) = 
||\B||_p = \left(\int_{\min(f_i)}^{\max(f_i)} |\B(\delta)|^p \,d\delta\right)^{1/p}
\end{equation}
This \(L^p\) norm then is closely related to computing
 total lifetime in terms of the information it summarises 
\citep{rieck2020topological}.  
To compare multiple Betti curves, \(\B_1\) and \(\B_2\), we naturally extend this notion and use the \(L^p\) distance between Betti curves, that is 
\begin{equation}
    L^p(\B_1, \B_2) = 
||\B_2 - \B_1||_p = \left(\int_{\min(f_i)}^{\max(f_i)} |\B_1(\delta)-\B_2(\delta)|^p \,d\delta\right)^{1/p}.
\end{equation}

\paragraph{Persistence Landscapes}

A frequently used tool for statistical analyses are persistence landscapes, which have been proven to follow a central limit theorem, even enabling the application of standard parametric tools \citep{bubenik2015statistical}.
First, the rank functions are defined for each persistence pair, \((p_i, q_i) \in \PD\), given as 
\begin{equation}  
\lambda_{(p_i, q_i)}(\delta) = 
\begin{cases}
 \delta - p_i & \text{ if } p_i < \delta \leq \frac{p_i + q_i}{2}\\
 - \delta + q_i & \text{ if } \frac{p_i + q_i}{2} < \delta < q_i\\
 0  & \text{ otherwise}\\
\end{cases}
\end{equation}
and the \(k\)-th persistence landscapes function is then given by \begin{equation}
    \Lambda(k,\delta) = \text{kmax}\{\lambda_{(p_i, q_i)}(\delta)\}_{i \in I}\end{equation} where kmax is the function giving the k-th largest value of a set. The persistence landscape then is the collection of these functions. Further, we can define the $L^p$ norm for persistence landscapes as:

\begin{equation}
L^p(\Lambda) = 
||\Lambda||_p = \sum_{k=1}^{\infty}\left(\int_{\min(f_i)}^{\max(f_i)} |\Lambda(k,\delta)|^p \,d\delta\right)^{1/p}\end{equation}

where the vertex feature \(f_i\) associated with \(x_i \in X\) determines the minimum and maximum value of our filtration parameter $\delta$.

\section{
Methods}\label{sec11}

\subsection{Computing Persistent Homology from Spatial Omics Data} 

Now, to develop our spatially variable gene detection method, we 
define spatial graphs from omics data 
in the manner that best matches 
the underlying sequencing technology \citep{tian_expanding_2022}. 
Amongst these technologies, spatial measurements are collected across varying spatial resolutions. 
For example, for data with continuous spatial coordinates, such as high resolution measurements of single cells, we compute the Delaunay triangulation to find an appropriate triangular covering of the space \citep{palla2022squidpy}. 
In comparison, Visium, one of the most popular sequencing methods, measures spots on a hexagonal grid with spots having diameters of $55 \mu m$ and centres being a fixed distance of $100\mu m$ apart \citep{moses2022museum}. Motivated by this, we choose a spatial graph connecting each grid point to its six directly adjacent neighbouring spots 
for further analysis. 
For both of these graph choices we then apply simplicial-complex based filtrations as explained in \cref{sec:ph}. 
Lastly, for data measured on rectangular grids, we connect each spot to its four adjacent neighbours and apply cubical persistence \citep{wagner2011efficient} as referenced in \cref{sec:ph} to capture the rectangular nature of these grid cells. 

Throughout this work we focus on computing zero-
dimensional persistent homology, which allows us to investigate the spatial patterns given by connected components. 
In particular, we are interested in encoding peaks in spatial gene expression and therefore construct superlevel-set filtrations using gene expression as the vertex feature for constructing the filtrations as described in \cref{sec:ph}. Then, the resulting persistence diagrams encode key information on spatial clustering behaviour of gene expression values across the tissue sample. Here, we choose to use a superlevel-set approach rather than a sublevel-set filtration because we reason that spatial maxima will be more suitable for revealing spatial trends than spatial minima for these data. This choice is informed by the fact that omics data is 
affected by technical noise and sparsity.  
Indeed, dropout is a well-known measurement error describing the common phenomena of not observing a transcript even though it is present. In practice, this effect leads to a large proportion of excess zeroes in gene expression values \citep{heumos2023best}, which could negatively affect the interpretation of local minima for our purpose. We therefore argue that spatial peaks will be more robust to such sparsity.

For each gene whose spatial pattern we want to assess, we then compute persistence diagrams and persistence summaries as detailed in \cref{sec:ph}.
By convention, we bound the death value of each persistence pair by the minimum filtration value. That is, rather than allowing one persistence feature to achieve an infinite death value, we bound it by the lowest gene expression value observed in the data. 
Note that our construction is similar to \citet{byers_detecting_2023}, but we further extend their methods to also consider cubical persistence for rectangular grids and investigate the use of functional summaries for spatial dependence testing as detailed below.

\subsection{One-Sample Permutation Testing for Spatial Dependence}
\label{sec:spatial_test}
We use randomized permutation testing to test for the following hypotheses:
\begin{compactitem}
    \item \(H_0\): Expression values are randomly distributed in space.
    \item \(H_1\): Expression values are not spatially random.
\end{compactitem}
We thus retain the original gene expression values as well as  the  spatial graph and randomly permute which expression value is assigned to each location to simulate spatial randomness, i.e. the lack of any spatial pattern. Then, significance is assessed as follows:

\begin{equation}
    \text{p-value} = \frac{\ones (||\widebar{\B_0} - \B_i||_p \leq ||\widebar{\B_0} - \B_{\text{obs}}||_p) + 1}{n_{\text{perm}} + 1} \end{equation} where \(\ones\) is the indicator function summing up the number of times the \(L^p\) norm, \(||\cdot||_p\), between the observed Betti curve \(\B_{\text{obs}}\) and the empirical mean Betti curve under the null model $\widebar{\B_0}$ is more extreme than the difference between each individual Betti curve under the null model and their overall mean. 

Note that we compute the mean Betti curve given the null hypothesis, \(\widebar{\B_0}\), as the mean Betti curve across all \(n_{\text{perm}}\) permutations as well as the observed Betti curve, while also adding 1 to both the numerator and denominator. 
This bias correction is included to avoid zero p-values and bias the test towards more conservative estimates by including this pseudo-count \citep{phipson2010permutation}. 
We thus conduct a two-sided, one-sample hypothesis test for each spatial feature. Whenever multiple genes are compared simultaneously, we further recommend applying a suitable multiple testing adjustment, such as  
Benjamini-Hochberg 
procedure to control the false discovery rate \citep{benjamini1995controlling}, which we will apply throughout our experiments. Across our experiments, the number of permutations is set to \(1000\), giving a practical compromise with regards to computational efficiency considerations. 

Beyond statistical reasoning, there also exist 
more deterministic approaches to spatially variable gene discovery. 
Here, the goal is to either rank features by the strength of their spatial dependence or to find a predefined number of the top most spatially variable features. 
In practice, these approaches are key for both explanatory analysis and for feature selection as a first step to uncover and understand spatial patterns in data \citep{andersson2021sepal, chen2024evaluating}. 
 For further analysis, we thus choose to rank spatially expressed features by 
their empirical p-values, following similar statistical approaches~\citep{li2023benchmarking}.

\subsection{Alternative Spatial Variability Detection Methods}
\label{sec:alternative}

In order to evaluate the performance of our spatial testing approach, we compare our methods with three standard methods for detecting spatial dependence. 

\paragraph{Moran's I:} Moran's I is one of the oldest and most widely-used measures for spatial autocorrelation. It is known to be a suitable baseline for SVG benchmarking \citep{li2023benchmarking} and can be computed as 
\begin{equation}
    I=\frac{n}{
    \sum_{i,j} w_{i,j}} 
    \frac{ \sum_{i=1}^{n}\sum_{j=1}^{n} w_{i,j} (f_i - \bar{f}) (f_j - \bar{f})}{
    \sum_{i=1}^{n} (f_i - \bar{f})^2
    }.
\end{equation}
Here, \(n\) corresponds to the number of observations, i.e. the number of vertices in the spatial graph \(G=(X,E)\). The feature \(f_i=f(x_i)\) corresponds to our gene expression measurement for observation \(x_i\in X\) and  \(\bar{f} = \frac{1}{n}\sum_{i=1}^n f_i\) is the mean feature value across observations. 
We use 
spatial adjacencies to define the weights \(w_{i,j}\), so that \(w_{i,j} = 1\) if \((x_i, x_j) \in E\) and \(w_{i,j} = 0\) otherwise. The value of Moran's I is directly interpretable, with $-1$ indicating negative autocorrelation, $0$ corresponding to spatial randomness, and $1$ indicating positive autocorrelation.
Further, for statistical analysis we use Moran's I in the same randomised permutation testing framework as the previously introduced topological summaries explained in \cref{sec:spatial_test}.

\paragraph{\texttt{sepal}: } \texttt{sepal} simulates diffusion processes and models a stochastic process that tracks the spread of a feature on the spatial domain over time as it tends towards randomness. Diffusion is simulated on either a rectangular or hexagonal grid and \texttt{sepal} then measures diffusion times, which are used to assess spatial patterns and measure how long it takes the diffusion process to converge. The higher the diffusion time, the stronger the spatial signal. Note that this method does not conduct any statistical testing and thus does not provide p-values. Instead, it provides a ranking of genes by their diffusion times. 
Moreover, \texttt{sepal} is only applicable to hexagonal or rectangular grids. Hence, whenever data is given with unstructured coordinates, we follow the implementation by \citet{andersson2021sepal} to convert spatial measurements  to a rectangular grid. 

\paragraph{\texttt{SpatialDE}:} Finally, \texttt{SpatialDE} applies spatial covariance testing via Gaussian process regression, a class of models originating from geostatistics. In a nutshell, gene expression is modelled via a multivariate normal model that includes both a non-spatial and a spatial covariance term.  
The model fitted with spatial covariance is then compared to a null model of spatial independence without this spatial component. 
Likelihood testing is used and p-values are computed analytically using the \(\chi^2\)-distribution with one degree of freedom. As a multiple testing correction, we follow \citet{svensson_spatialde_2018} and report q-values, which specifically control for the positive false discovery rate \citep{storey2003statistical}. Because \texttt{SpatialDE} assumes normally distributed residual noise, we follow the preprocessing steps by \citet{svensson_spatialde_2018}. First, Anscombe’s transformation, a variance stabilising method for negative binomial data, is applied. Second, log total count values are regressed out 
to achieve independence from the total counts per spatial location.

\subsection{Datasets and Preprocessing}
\label{sec:omics_data}
%
We use three real transcriptomics datasets for our investigation. First, we analyse spatial data of a mouse olfactory bulb (referred to as MOB, replicate 1) and a breast cancer tumour sample (breast, replicate 1) from \citet{staahl2016visualization}, standard datasets commonly used to test spatial variability detection methods. Both have been sequenced using 1k spatial transcriptomics arrays, which measure spots on rectangular grids. Each spot has a diameter of $100 \mu m$ and encompasses 10--100 cells. 
Further, we study one sample from \citet{kuppe_spatial_2022} (sample IZP10), which consists of spatial measurements of spots on a hexagonal grid. This sample has been taken from tissues with necrotic areas extracted from the heart of a patient that suffered from a heart attack. 

Prior to analysis, preprocessing and quality control of omics data are crucial steps for accounting for technical errors as well as filtering out noise. Indeed, normalisation is a key step in spatial omics data analysis as many downstream tasks, including the detection of spatially variable features, rely on adequate preprocessing choices to account for technical effects \citep{heumos2023best}.
Following best practices \citep{heumos2023best} and alternative spatial variability studies \citep{edsgard2018identification}, we decide on the following preprocessing strategy and
select all genes that 
   have a minimum total expression count of $10$ and 
 are present in at least 1\% of locations. 
Further, we remove all cells with fewer than $10$ counts. These steps are done to avoid analysing cells or genes for which hardly any biological signals have been detected. Finally, 
we exclude all ribosomal and mitochondrial genes from analysis as these indicate technical artifacts and are not related to the mRNA, i.e.\ protein-coding genes that are the focus of transcriptomics data analysis \citep{palla2022squidpy}. 
For each real dataset, we then identify the $500$ most variable genes using \texttt{\(\texttt{seurat}_{\texttt{v3}}\)} 
for further analysis. That is, we select the set of genes that vary most notably across cells \citep{heumos2023best}. Our goal will then be to detect whether these genes that differ noticeably between dislocated cells also show distinct spatial patterns. Note that for the simulated data we skip these quality control steps as we want to compare all SVG detection methods across all simulated features. 

Further, as conventionally done in spatial omics data analysis, we decide to preprocess the expression values, i.e. gene expression count data, 
in the manner best fitting the SVG detection method \citep{li2023benchmarking, chen2024evaluating}. For \texttt{SpatialDE} we apply the normalisation strategy preferred for this method, as detailed in \cref{sec:alternative} \citep{svensson_spatialde_2018}.
For \texttt{sepal}, we apply the recommended shifted logarithm normalisation by applying the transformation \(\log(f+2)\) to each feature $f$. This transformation is applied to 
stabilize the variance across the dataset to counteract variable sampling effects, such as overdispersion, and reduce the skewedness of the feature's distribution. The pseudo-count of $2$ is added to enable the log-transformation also for zero-values counts. To ensure comparability across spatial variability tests, we similarly choose to compute our persistence-based test as well as Moran's I from the same $\log(f+2)$-transformed features. 
Note however, that this preprocessing strategy is not the only viable choice and our method can be flexibly applied across varying preprocessing choices \citep{ahlmann2023comparison}.
In general, given that our persistence-based methods directly scale through each gene's expression values, we highly recommend to apply our framework to transformed and normalised features, but we leave the choice of the best data-dependent transformation to further investigation \citep{ahlmann2023comparison}.

\subsection{Synthetic Data Generation}
\label{sec:synt}

\begin{figure}[tb]
\centering
\includegraphics[width=1\textwidth]{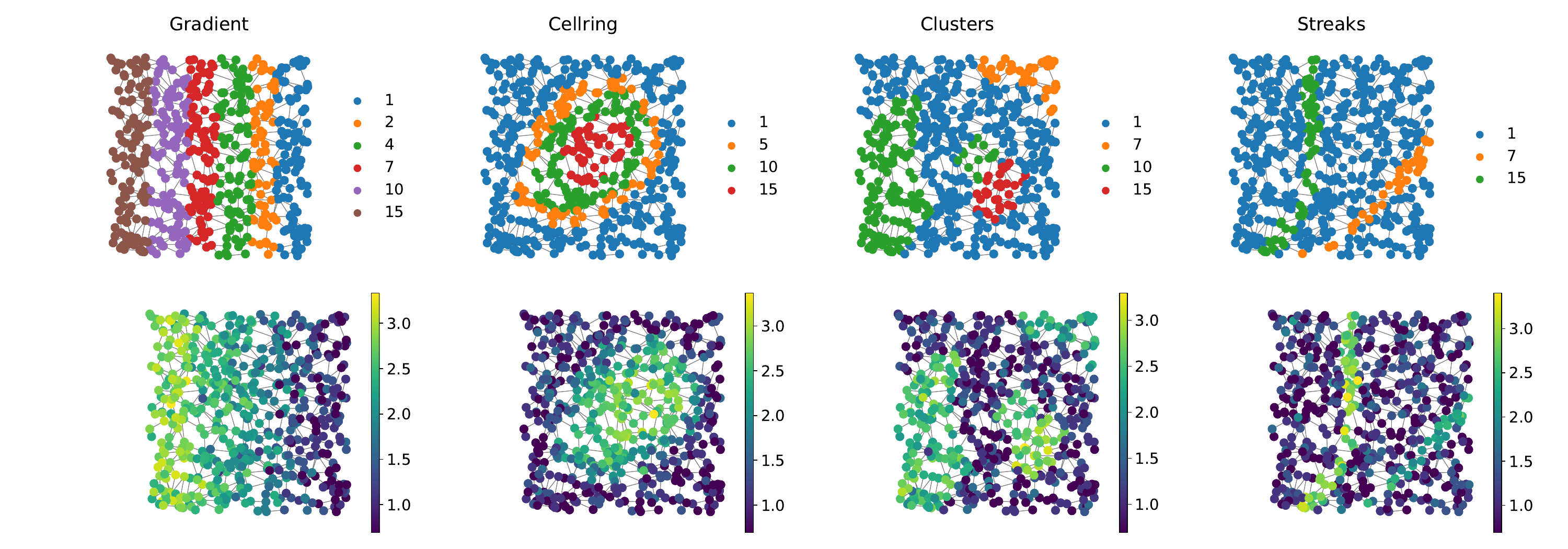}
\caption{Simulated spatial patterns with varying effect sizes for each spatial domain as indicated in the legend (row 1). Examples of simulated genes with these spatial signals (row 2). The simulated patterns are showing a gradient, a cellring, a set of clusters, and two lines from left to right. 
}\label{fig:graphs_toy}
\end{figure}

We create in-silico examples of spatial transcriptomics data using SRTsim \citep{zhu2023srtsim}, an R package for the simulation of spatial patterns. In particular, we randomly sample spatial locations on a square domain and then simulate count data using (zero-inflated) Poisson or negative binomial distributions. 
The choice between these range of distributions of count data is motivated by the fact that the observed distribution and nature of transcriptomics data is highly debated and varies between sequencing technologies and datasets. 
Zero-inflation aims to account for the excess zeroes often encountered in sparsely sampled gene expression data due to a high dropout probability. 
Across our simulations we thus consider zero-inflation 
to be a potential source of noise and assess the robustness of the proposed spatially variable gene detection methods against these technical errors. 
SRTsim allows us to choose the following settings for simulations. Globally, we set the mean parameter (\(\mu = 1\)), the dispersion parameter (\(r\)), and the zero-proportion parameter (\(z\)). 
We then define \(N\) spatial domains corresponding to distinct spatial shapes that are assigned an effect size \(e_i\) resulting in an \(e_i\)-fold increase in mean for the specified spatial domain \(i \in {0, 1, 2, .., N}\). The effect size of the background domain is set to \(e_0=1\). Count data is then sampled separately for each domain with the thus specified parameters and assigned randomly to locations within the shape. 
We then simulate four distinct spatial patterns: a gradient, two lines, a set of clusters, and a cellring illustrated in \cref{fig:graphs_toy} with the annotated effect sizes. These spatial signals are to mimic distinct spatial tissue structures which could occur in real datasets \citep{zhu2023srtsim, edsgard2018identification, andersson2021sepal}. We also simulate genes without spatial signal, whose gene expression values are sampled randomly across the spatial locations.

\section{Results}\label{sec:results}

Across our experiments we consider two main goals: (i) using significance testing to determine spatially dependent features, and (ii) ranking features by their spatial dependence and identifying a fixed number of the top most spatially variable genes. 
Throughout our empirical experiments we make the following observations:
\begin{compactitem}
    \item PH-based approaches perform well at detecting SVGs in terms of the area under the precision recall curve. They are more robust to zero-inflation, i.e. high proportions of excess zeroes in gene expression counts, than Moran's I. 
    \item PH-based significance testing leads to lower sensitivity but higher specificity identifying a smaller set of SVGs at a fixed $0.05$ significance threshold. 
    \item PH-based approaches are superior to alternatives at detecting a predefined number of the top most spatially variable genes. 
    \item PH-based approaches capture orthogonal information to alternative methods.
    \item PH-based spatial dependence tests are less correlated to the total sum of feature values than Moran's I.
\end{compactitem}
We further explore the use of our PH-based approach to detect SVGs for three real spatial omics datasets, for which we plot and examine the top spatially variable features. 
We thus demonstrate that persistence-based methods, and Betti curves in particular, offer valid alternatives to existing SVG methods, such as Moran's I. 
\cref{tab_results} summarises the performance of each SVG detection methods across all our experiments, which we will now elaborate on in more detail.

\begin{table}[h!]
\caption{Summary of the ranking of each SVG detection method by their performance across experiments. The lower the rank the better. Persistence-based permutation tests achieve high median ranks.}
\label{tab_results}
\footnotesize
\begin{tabular}{@{}ll>{\centering\arraybackslash}m{1.4cm}>{\centering\arraybackslash}m{1.4cm}>{\centering\arraybackslash}m{1.4cm}>{\centering\arraybackslash}m{1.4cm}>{\centering\arraybackslash}m{1.4cm}>{\centering\arraybackslash}m{1.4cm}@{}}
\toprule
& & \multicolumn{6}{c}{\textbf{Ranking of SVG Methods}} \\ \midrule
\textbf{Experiment} & \textbf{Dataset} & \textbf{Betti}  \textbf{Curves} & \shortstack{\textbf{Pers.}\\\textbf{Landscapes}} & \shortstack{\textbf{Total} \\\textbf{Pers.}} & \textbf{SpatialDE} & \textbf{Moran's I} & \textbf{sepal} \\ \midrule

\multirow{4}{*}{\shortstack{AUPRC \\(Fig. 3, Sec. 5.1.1\\ to 5.1.5)}}
  & Gradient & \rankcolor{2} & \rankcolor{1} & \rankcolor{2} & \rankcolor{5} & \rankcolor{4} & \rankcolor{6} \\
  & Cellring & \rankcolor{3} & \rankcolor{2} & \rankcolor{3} & \rankcolor{5} & \rankcolor{6} & \rankcolor{1} \\
  & Clusters & \rankcolor{2} & \rankcolor{1} & \rankcolor{2} & \rankcolor{6} & \rankcolor{4} & \rankcolor{5} \\
  & Streaks & \rankcolor{3} & \rankcolor{2} & \rankcolor{3} & \rankcolor{6} & \rankcolor{3} & \rankcolor{1} \\
\midrule

\multirow{4}{*}{\shortstack{Sensitivity\\(Fig. 4, Sec. 5.1.6)}}
  & Gradient & \rankcolor{3} & \rankcolor{4} & \rankcolor{1} & \rankcolor{5} & \rankcolor{1} & / \\
  & Cellring & \rankcolor{3} & \rankcolor{4} & \rankcolor{1} & \rankcolor{5} & \rankcolor{1} & / \\
  & Clusters & \rankcolor{3} & \rankcolor{4} & \rankcolor{1} & \rankcolor{5} & \rankcolor{1} & / \\
  & Streaks & \rankcolor{3} & \rankcolor{3} & \rankcolor{1} & \rankcolor{5} & \rankcolor{1} & / \\
\midrule

\multirow{4}{*}{\shortstack{Specificity\\(Fig. 4, Sec. 5.1.6)}}
  & Gradient & \rankcolor{2} & \rankcolor{1} & \rankcolor{2} & \rankcolor{4} & \rankcolor{5} & / \\
  & Cellring & \rankcolor{1} & \rankcolor{1} & \rankcolor{1} & \rankcolor{4} & \rankcolor{5} & / \\
  & Clusters & \rankcolor{2} & \rankcolor{1} & \rankcolor{2} & \rankcolor{4} & \rankcolor{5} & / \\
  & Streaks & \rankcolor{1} & \rankcolor{1} & \rankcolor{1} & \rankcolor{4} & \rankcolor{5} & / \\
\midrule

\multirow{4}{*}{\shortstack{Correlation with\\ total  counts \\ (Fig. 6, Sec. 5.2.1)}}
  & Gradient & \rankcolor{2} & \rankcolor{1} & \rankcolor{2} & \rankcolor{4} & \rankcolor{5} & / \\
  & Cellring & \rankcolor{1} & \rankcolor{1} & \rankcolor{1} & \rankcolor{4} & \rankcolor{5} & / \\
  & Clusters & \rankcolor{2} & \rankcolor{1} & \rankcolor{2} & \rankcolor{4} & \rankcolor{5} & / \\
  & Streaks & \rankcolor{1} & \rankcolor{1} & \rankcolor{1} & \rankcolor{4} & \rankcolor{5} & / \\
\midrule

\multirow{4}{*}{\shortstack{Top genes\\ identified \\ (Fig. 6, Sec. 5.2.2)}}
  & Gradient & \rankcolor{2} & \rankcolor{2} & \rankcolor{2} & \rankcolor{2} & \rankcolor{6} & \rankcolor{1} \\
  & Cellring & \rankcolor{1} & \rankcolor{2} & \rankcolor{1} & \rankcolor{1} & \rankcolor{6} & \rankcolor{5} \\
  & Clusters & \rankcolor{1} & \rankcolor{2} & \rankcolor{1} & \rankcolor{1} & \rankcolor{6} & \rankcolor{5} \\
  & Streaks & \rankcolor{1} & \rankcolor{2} & \rankcolor{1} & \rankcolor{1} & \rankcolor{6} & \rankcolor{5} \\
\midrule

\multirow{3}{*}{\shortstack{Number of SVGs\\ (Fig. 7, Sec. 5.3.2)}}
  & MOB & \rankcolor{2} & \rankcolor{4} & \rankcolor{3} & \rankcolor{4} & \rankcolor{1} & / \\
  & Breast & \rankcolor{2} & \rankcolor{5} & \rankcolor{5} & \rankcolor{3} & \rankcolor{1} & / \\
  & Heart & \rankcolor{1} & \rankcolor{4} & \rankcolor{2} & \rankcolor{5} & \rankcolor{3} & / \\
\midrule

\multirow{3}{*}{\shortstack{Correlation with\\total counts \\ (Fig. 7, Sec. 5.3.3)}}
  & MOB & \rankcolor{5} & \rankcolor{3} & \rankcolor{4} & \rankcolor{1} & \rankcolor{6} & \rankcolor{2} \\
  & Breast & \rankcolor{3} & \rankcolor{2} & \rankcolor{1} & \rankcolor{4} & \rankcolor{6} & \rankcolor{5} \\
  & Heart & \rankcolor{5} & \rankcolor{1} & \rankcolor{2} & \rankcolor{4} & \rankcolor{3} & \rankcolor{6} \\
\midrule

\textbf{Median Rank} & & \textbf{\rankcolor{2}} & \textbf{\rankcolor{2}} & \textbf{\rankcolor{2}} & \textbf{\rankcolor{4}} & \textbf{\rankcolor{4}} & \textbf{\rankcolor{5}} \\
\bottomrule
\end{tabular}
\end{table}


\subsection{Simulation Study}

\label{sec:sim}

\subsubsection{PH-based approaches are robust to zero-inflation}

\begin{figure}[tb]
\centering
\includegraphics[width=1\textwidth]{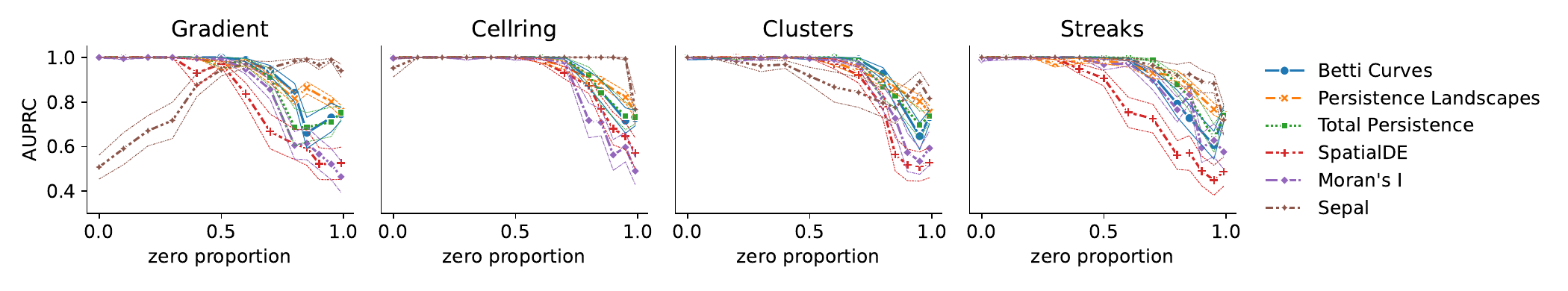}
\caption{Area under the precision-recall curve (AUPRC) across varying degrees of zero inflation for different spatial variability detection methods. Slim lines indicate the standard deviation across 1000 bootstrap re-samples.}\label{fig:auprc_toy}
\end{figure}

We simulate spatial data as described in \cref{sec:synt} and first vary the zero-inflation parameter \(z\) from \(0.1\) to \(0.99\), keeping the dispersion fixed at \(0.3\). For each shape and parameter choice, we then simulate $50$ genes that possess the spatial signal 
and \(50\) genes that are randomly sampled without any spatial effect. 
Then, across each degree of zero-inflation, we summarise the performance of each statistical test and scoring method in the following manner: 
\cref{fig:auprc_toy} 
reports the area under the precision-recall curve (AUPRC) as a function of the degree of zero-inflation for all methods. 
This comparison is chosen because it allows us to compare all methods, also the ranking-based \texttt{sepal} method, as a single summary statistic per degree of zero-inflation. 
An AUPRC of \(1\) indicates perfect performance, whereas random guessing is expected to result in an AUPRC of \(0.5\), and \(0\) is the worst possible value. 

\subsubsection{Persistence-based testing performs well at detecting SVGs}
We observe that all persistence-based methods achieve high AUPRC values very close to $1$ until around $70\%$ of zero-inflation across all patterns, as summarised in \cref{fig:auprc_toy}. 
This is followed by a decrease in AUPRC, which remains above \(0.6\), even for \(99\%\) excess zeros. Overall, these trends show the robustness and expressivity of persistent homology as a summary of spatial patterns, even in the presence of high zero-inflation. 

\subsubsection{Persistence-based methods perform similarly to one another}
We further observe that both Betti curves and total persistence 
follow very similar trends in \cref{fig:auprc_toy}, with Betti curves performing better on the gradient pattern and total persistence showing higher scores on the streaks pattern.
This finding highlights that both approaches are alike in terms of the information they are summarising.
Persistent landscapes meanwhile achieve consistently higher AUPRC values above $0.7$ for higher degrees of zero-inflation across datasets and thus capture trends in spatial variability in a more robust manner for high degrees of dropout.
For this experiment, we thus observe that persistence landscapes summarise different aspects of persistent homology than alternative summaries. This slight difference in performance reported in \cref{fig:auprc_toy} arises because persistence landscape functions \citep{bubenik2015statistical} differ in the way they are constructed from a given persistence diagram compared to Betti curves, which are more alike to total persistence \citep{rieck2020topological}. 

\subsubsection{Persistence-based methods outperform Moran's I on zero-inflated data}
Our results in \cref{fig:auprc_toy} indicate that persistent homology is more expressive at quantifying spatial signals in simulated expression values under the presence of zero-inflation than Moran's I. 
That is, we observe that by tracking the evolution of connected components via superlevel-set filtrations 
on spatial graphs, we get summary statistics with higher discriminative performance than Moran's I, which summarises spatial autocorrelation based on summarising local neighbourhoods and is the most commonly used baseline measure for spatial variability. Indeed, via this comparison in terms of AUPRC, we assess that persistence-based summaries offer a valid alternative to established tools from spatial statistics.

\subsubsection{Persistence detects spatial variability across all simulations} 
We note that our proposed persistence-based tests do well at detecting spatial variability, in particular for low levels of zero-inflation, as would be expected. However, in comparison we see that some alternative methods do not follow the same clear trend across datasets. \texttt{SpatialDE} starts decreasing in terms of AUPRC at comparably lower proportions of zero inflation. This difference is especially pronounced for the streaks pattern, where \texttt{SpatialDE} performs notably worse than all alternative methods after \(60\%\) zero-inflation. Potentially this is due to this simulation only showing two relatively thin lines (as plotted in \cref{fig:graphs_toy}) that are harder for \texttt{SpatialDE} to detect as spatially variable under the presence of excess zeroes. 
Meanwhile, 
\texttt{sepal} performs very well for higher values of \(z\) between \(0.6\) and \(0.8\) across datasets, showing higher AUPRCs than persistence landscapes. However, \texttt{sepal} fails at detecting the gradient pattern at low degrees of zero-inflation. For example, \texttt{sepal} receiving an AUPRC of only \(0.6\) at \(z=0.1\) highlights that diffusion times are not suitable for detecting this smooth trend in spatial variability. We thus demonstrate disadvantages of both \texttt{sepal} and \texttt{SpatialDE} compared to our persistence-based approach. 
We conclude that PH-based methods achieve overwhelmingly higher or similar AUPRCs to the baseline method Moran's I, highlighting their utility as robust SVG detection methods.

\begin{figure}[tb]
\includegraphics[width=0.85\textwidth]{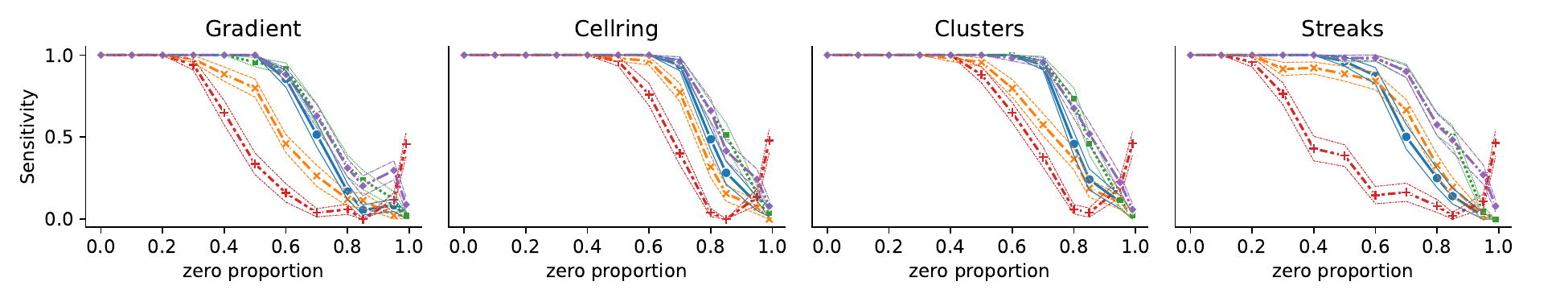}

\includegraphics[width=1\textwidth]{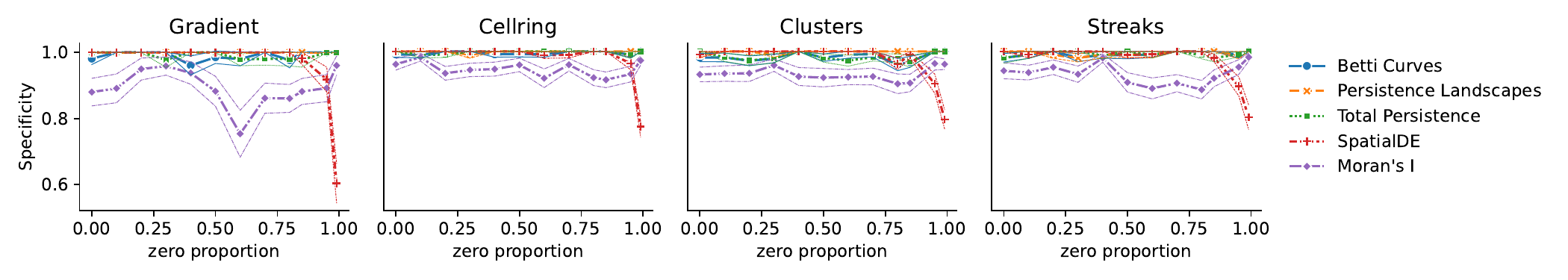}
\caption{Sensitivity and specificity using a fixed 0.05 cut-off after multiple testing correction. 
Slim lines indicate the standard deviation across 1000 bootstrap re-samples.}\label{fig:sens_toy}
\end{figure}

\subsubsection{PH-based methods are more specific than Moran's I}
Extending our observations beyond AUPRC, we next decide to view spatially variable gene detection in the context of hypothesis testing. \cref{fig:sens_toy} compares the performance of each 
statistical test in terms of specificity and sensitivity at a fixed \(0.05\) significance 
threshold for corrected p-values. 
In agreement with results from \citet{byers_detecting_2023}, we observe that 
at this fixed threshold, total persistence is either similarly or slightly less powerful, i.e. less sensitive than Moran's I. This is followed by
Betti curves, which detect a lower proportion of true positives, in particular on the streaks pattern, as well as persistence landscapes which give even more conservative estimate. However, we also observe that all persistence-based approaches show very high specificity across examples, whereas Moran's I demonstrates the highest false positive rate of around \(5\%\) to \(25\%\) even after p-value correction, which is undesirable in practice when the set of spatially variable genes should be correctly identified. 

Our findings thus agree with \citet{byers_detecting_2023}, who find that persistence is less sensitive than Moran's I. 
However, \citet{byers_detecting_2023} further claim that persistence cannot replace Moran's I, a finding we do not fully support. Arguably, specificity as well as sensitivity are important for interpreting spatial dependency tests in practice. Indeed, we observe that at a default \(0.05\) significance cutoff, corrected p-values from PH-based approaches achieve higher specificity than Moran's I, showing that PH-based results can be more useful for tasks that require a low false positive rate and benefit from a more conservative assessment. Given the high AUPRC achieved by PH-based methods, we further assess that choosing a higher significance threshold could be a suitable adjustment to achieve higher sensitivity and identify a larger set of spatially variable features while retaining superior specificity.

\subsubsection{Agreement between methods}

\begin{figure}[tb]
\centering
\includegraphics[width=1\textwidth]{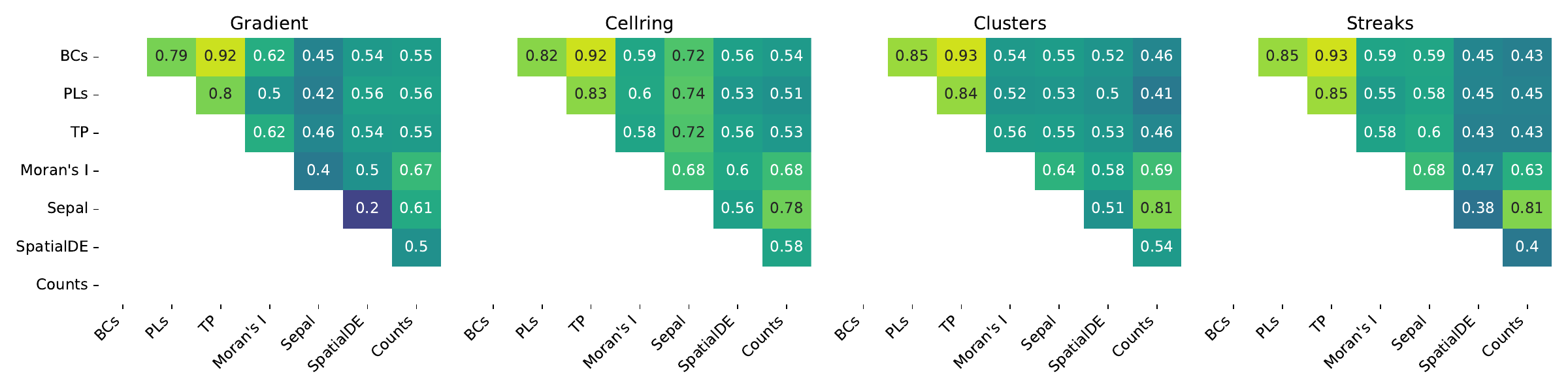}
\caption{Spearman correlation between the rankings given by different spatial variability detection methods across the four spatial patterns from \cref{fig:graphs_toy}. Generally, there is low to medium correlation between methods with the highest agreement being between total persistence and Betti curves. Persistence thus captures distinct characteristics in spatial variability not identified by other approaches.}\label{fig:corr_toy}
\end{figure}

Further, we compute the Spearman correlation between each method's ranking 
and report the results in \cref{fig:corr_toy}.
For persistence-based approaches, we rank genes by their
p-values, else we use the values of the scores as defined in \cref{sec:alternative} for \texttt{sepal} and Moran's I. 
From the results in \cref{fig:corr_toy}, we observe that total persistence and Betti curves perform most similarly across experiments with correlations between 0.64 and 0.82, which is likely driven by their common roots in accumulating persistence features \citep{rieck2020topological}. 
Further, we find that there is varying but overall moderate levels of agreement between persistence and Moran's I (\(\rho \leq 0.53\)), likely explained by the fact that both are summaries of the same spatial graph and understand spatial variability via modelling clustering behaviour, as summarised by connected components in PH, or autocorrelation, as summarised by differences in neighbourhood scores. 
Interestingly, while \texttt{sepal} follows a fundamentally different modelling approach, correlation between \texttt{sepal} and the other approaches 
exists at comparable levels to other entries in the correlation matrix and the \texttt{sepal} score is actually most correlated with Moran's I on the clusters dataset (\(\rho = 0.58\)). Overall, we thus find that persistence-based permutation testing captures complementary information not encoded by either Moran's I, \texttt{sepal}, or \texttt{SpatialDE}.

\subsection{PH-based approaches identify the top SVGs}

Next, we keep the zero proportion at \(z=0.1\) and the dispersion parameter at \(r=0.3\), following estimations by  \citet{zhu2023srtsim} from real data, but vary the effect size by taking \(\hat{e_i}~=~max(e_i / c, 1)\) for some \(c\) in \(\{ 6, 5, 4, 3, 2, 1\}\) for each domain \(i=1,2,...,N\) effectively varying the strength of the spatial pattern compared to the background domain, for which \(e_0=1\). For each parameter choice and spatial shape, we then simulate 50 features with spatial signals and 50 completely random patterns. 
We then investigate the identification of top spatially variable genes for different strengths of spatial effects.

\begin{figure}[tb]
\centering
\includegraphics[width=0.75\textwidth]{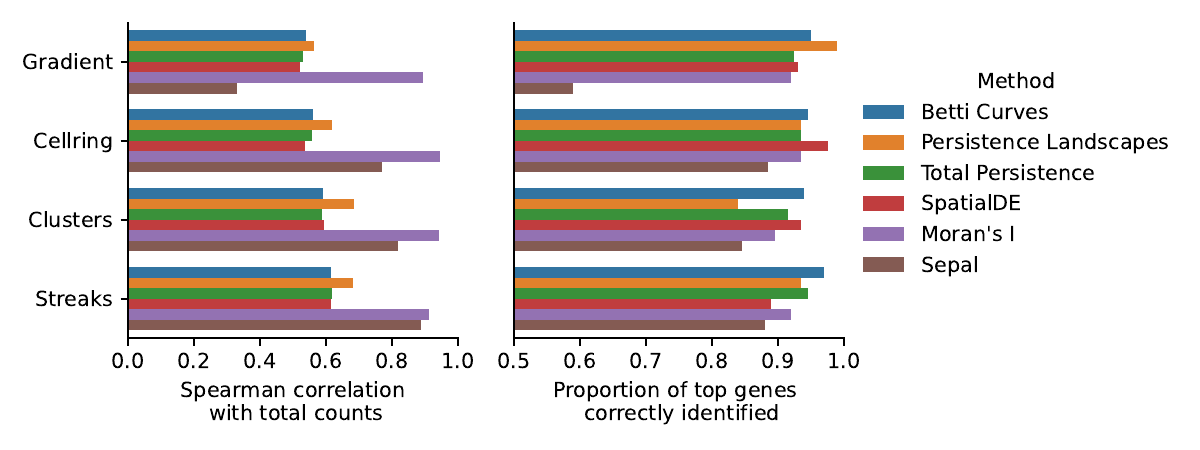}
\caption{Comparison between the correlation between total counts and spatial variability scores as well as the proportion of true positives in the set of top genes.}\label{fig:rank_toy}
\end{figure}

\subsubsection{Correlation with the sum of gene expression counts}
As reported in \cref{fig:rank_toy}, we explore the correlation between each spatial variability score and the total count per gene, i.e. the sum of feature values. 
We want an ideal SVG detection method to not just correspond to these total counts, but score genes based solely on the existence of spatial dependencies in gene expression. 
Across datasets, we see that Moran's I and \texttt{sepal} show the highest correspondence to the sum of feature values. In comparison,  persistence-based rankings achieve consistently lower Spearman correlation of $0.6$ to the number of total counts. Persistence landscapes reach slightly higher correlation than Betti curves and total persistence. Overall, this rank correlation is similar to the one achieved by \texttt{SpatialDE}. Therefore, the comparison highlights that persistence offers a satisfactory alternative to established SVG tests.

\subsubsection{PH-based approaches correctly identify the top SVGs}
Across methods, we then compare the ground truth 
given by the existence of spatial signal with each spatial variability score.  \cref{fig:rank_toy} shows the proportion of truly positive features detected by each method. Ideally, a suitable spatial variability detection method would show a high proportion of true positives amongst the top 300 genes (given that 300 out of 600 features show spatial patterns). 
Notably, Betti curves perform consistently well across examples correctly identifying more than \(90\%\) of top SVGs. This is followed by total persistence and persistence landscapes, which perform better on the gradient but worse on the cluster pattern. 
Interestingly, from these examples we see that while the proportion of genes correctly identified is high across  scenarios, there is no strong distinction between either persistence- or non-persistence-based methods based on these examples. However, \texttt{sepal} and Moran's I seem to perform slightly worse at identifying the most spatially variable features than persistence-based approaches, which together with their correlation to total counts indicates that persistence-based testing is preferable in this scenario. 

\begin{figure}[tb]
\centering
\includegraphics[width=0.4\textwidth]{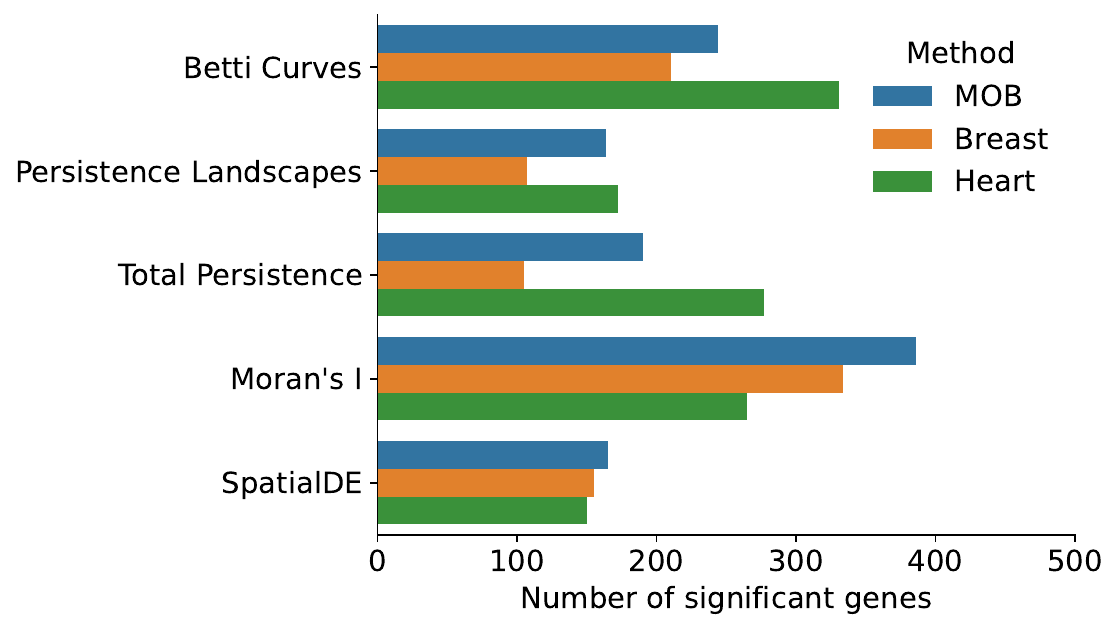}
\includegraphics[width=0.45\textwidth]{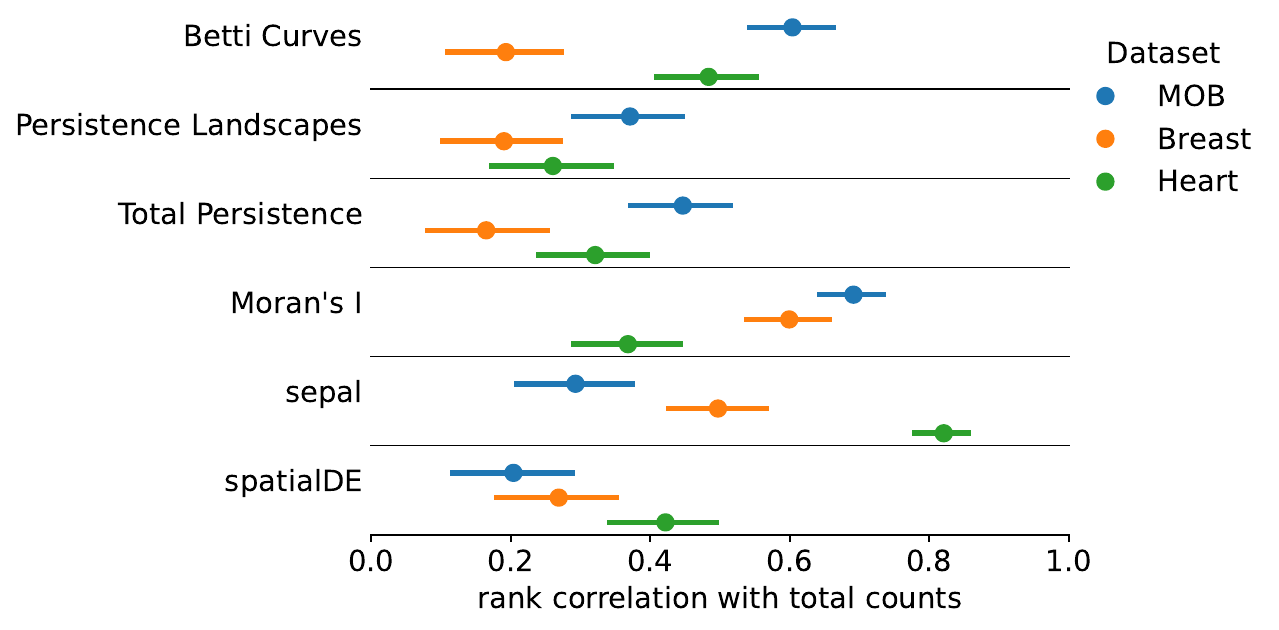} 
\caption{Number of spatially variable genes detected by each method (left) and the rank correlation between each method's score and the total number of counts per gene (right) with standard deviations computed across 1000 bootstrap re-samples. 
}\label{fig:sig_real}
\end{figure}
\begin{figure}[t]
\centering
\includegraphics[width=1\textwidth]{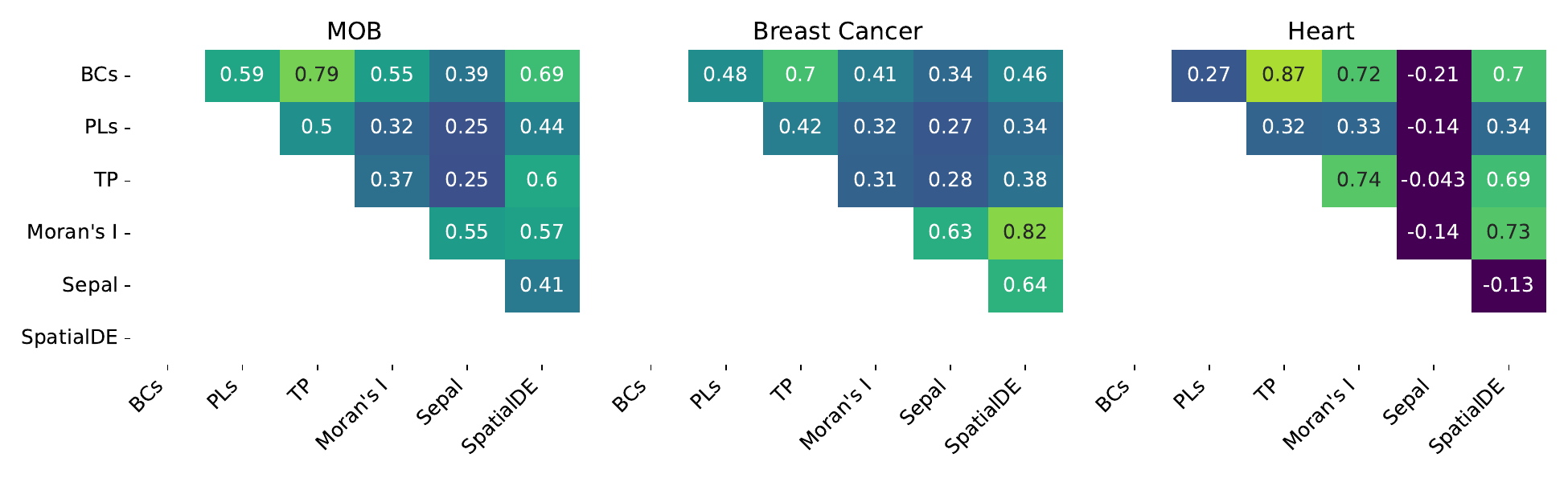}
\caption{Correlation between the ranking of all genes ranked by each method for the mouse olfactory bulb (left), the breast cancer slide (middle) and the Visium heart sample (right). 
}\label{fig:corr_real}
\end{figure}

\begin{figure}[!b]
\centering
\includegraphics[width=1\linewidth]{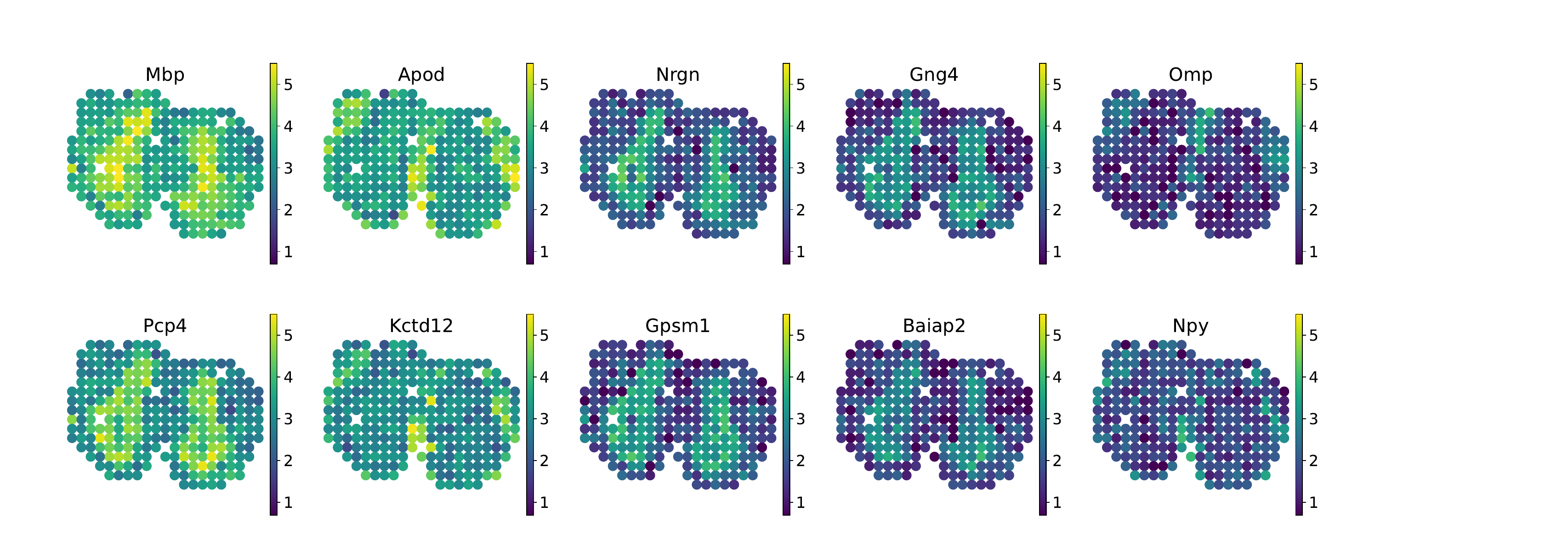}
\includegraphics[width=1\linewidth]{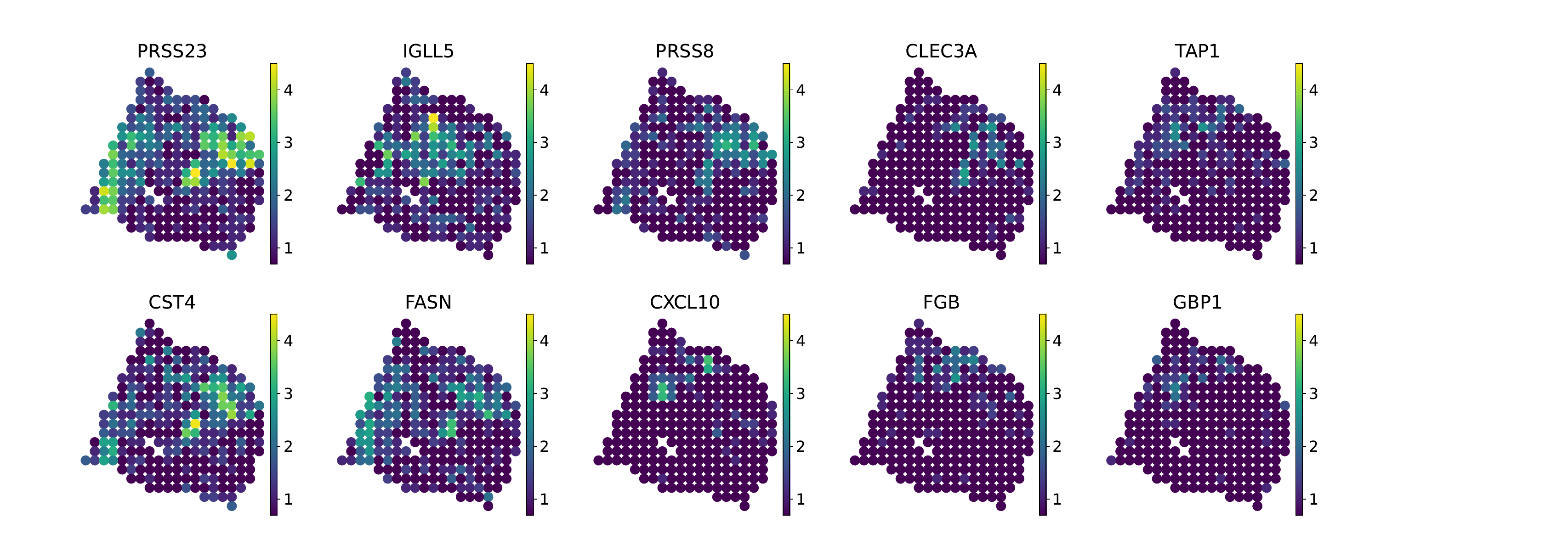}
\includegraphics[width=1\linewidth]{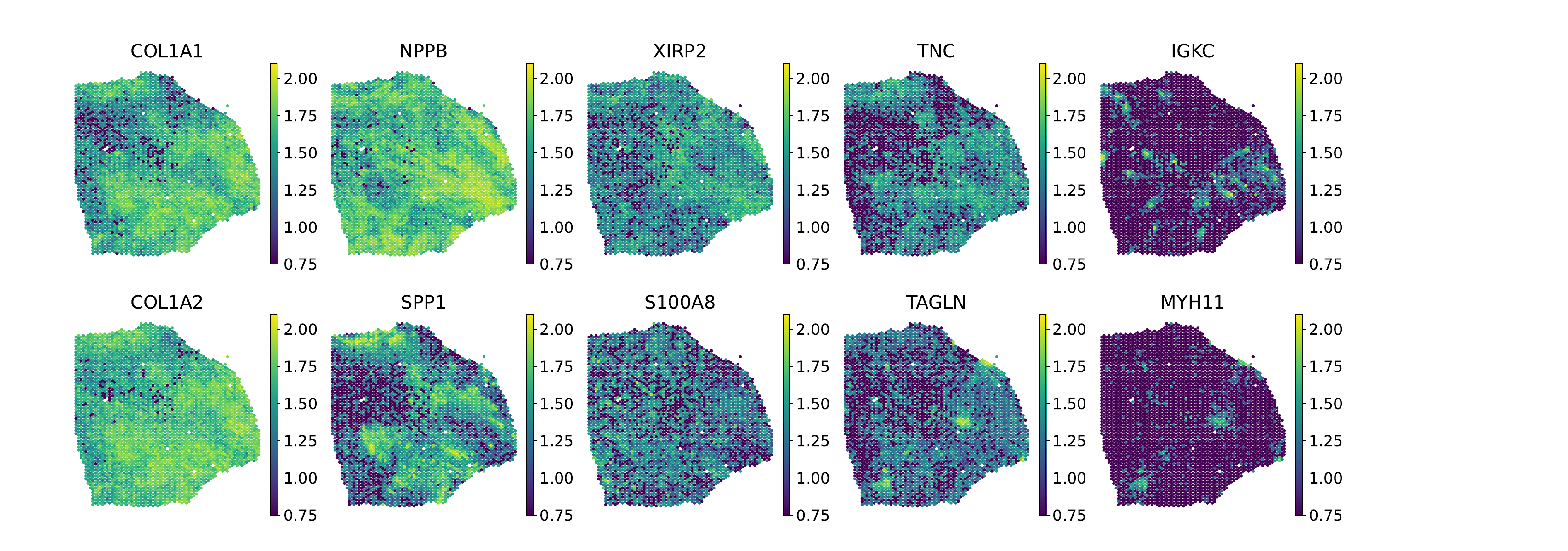}
\caption{Examples of the most spatially variable genes as identified by Betti curves for the mouse olfactory bulb (top), the breast cancer sample (middle), and the heart data (bottom). Each plot shows the spatial distribution of one gene's expression values preprocessed as detailed in \cref{sec:omics_data}. The name of each gene is plotted above the respective tissue sample.} 
\label{fig:examples_real}
\end{figure}

\subsection{Analysing Transcriptomics Data}
\label{data}
Having observed that persistence-based approaches give suitable spatial variability tests on simulated data, we next investigate their utility on real datasets of spatial tissue samples as detailed in \cref{sec:omics_data}.

\subsubsection{PH-based methods identify spatially variable genes}

First, we decide to focus on one of our proposed methods, namely Betti curves, which performed well across our simulated case study, and investigate spatially variable genes detected by this approach. \cref{fig:examples_real} shows representative examples of genes that are both the most highly variable and the most spatially variable features reaching the lowest possible p-values in terms of the spatial dependence test. We see that Betti curves identify both genes with low and high expression counts to be spatially variable. 
For the mouse olfactory bulb (MOB), we see that there is a clear pattern of genes being more highly expressed and clustered in the centre of this brain region, such as Mbp, Pcp4, Nrgn, or Gpsm1. What these genes have in common is that they relate to neural activity in this region of the mouse's brain \cite{li2024sgcast}. For the breast cancer sample, we see some more sparse patterns that could indicate disease-specific differences in tissue regions and immune response. Finally, for the heart sample, the genes COL1A1 and COL1A2 are found to be spatially variable, which are coding for collagen type~1, mainly found in connective tissue and related to wound healing. These genes have been found to be unregulated in tissue regions impacted by myocardial infarction \cite{van2008dysregulation} and are thus of biological relevance for understanding the spatial composition of this tissue sample. Overall, studying the set of spatially variable genes in more depth can thus reveal important insights into tissue biology and disease response by specifically focusing on transcripts that vary across tissue regions.

\subsubsection{PH-based methods identify fewer SVGs than Moran's I}
Across the 
real datasets considered for this study, we  report the number of spatially variable genes detected by each method. Results in \cref{fig:sig_real} then confirm that Moran's I reports the largest number of significantly spatially autocorrelated genes, calling more than $50\%$ of transcripts spatially variable. This is followed by Betti curves, which find more spatially variable patterns for the heart sample, but fewer for the mouse brain and breast cancer sample. Interestingly, across these real datasets, total persistence detects a low number of SVGs, which is comparable to persistence landscapes but  notably lower than Betti curves. This trend has not been observed across our simulations, indicating that the results of using Betti curves or total persistence are not necessarily similar.  
Overall, the reported numbers of SVGs support our findings from the simulation study in \cref{sec:sim} on the sensitivity and specificity of each approach. We conclude that persistence-based methods identify a smaller number of genes than Moran's I, but do so with a lower type-1 error rate.

\subsubsection{Persistence is robust to the sum of feature values}
Next, we investigate the dependence of each score on the total counts measured for each gene as reported in \cref{fig:sig_real}. Ideally, spatial dependence tests should be robust to whether features have high or low values overall. That is, a good method for detecting spatially variable genes should not be biased towards genes with high expression levels, but also detect spatial patterns in lowly expressed genes.  
We report the Spearman correlation with total counts as a sanity check. Amongst all methods, we see that Moran's I generally shows the highest correlation with total counts. In contrast, persistence-based methods are more robust to the sum of feature values. Finally, \texttt{sepal} is designed to quantify spatial variability independently of total counts, but we find that in some examples it actually correlates more with total counts than persistence-based methods.

\subsubsection{Agreement between methods}
Further, \cref{fig:corr_real} reports the rank correlations between the gene rankings reported by each spatially variable gene detection method. Overall, we once again see a high agreement between Betti curves and total persistence, with \(\rho \in [0.7, 0.79]\). This is followed by moderate correlation to persistence landscapes, Moran's I, and \texttt{SpatialDE}, with \(\rho \in [0.32, 0.69]\). Persistence landscapes rank genes in a notably different order than other persistence-based approaches, with \(\rho \in [0.42, 0.59]\). 
Overall, the least agreement is reported between persistence-based methods and \texttt{sepal}, with \(\rho \in [0.24, 0.39]\). Overall, our results point towards a relatively high disagreement between methods. Similar to the simulation study reported in \cref{fig:corr_toy}, we find that persistence summarises distinct information on spatial patterns not encoded by existing SVG detection methods.

\section{Discussion}
\label{sec:discussion}

Throughout our study, we proposed randomised permutation testing procedures for spatial variability in omics data via Betti curves or persistence landscapes. We \mbox{investigated} these tests in comparison to baseline methods Moran's I, \texttt{sepal}, \texttt{SpatialDE}, and total persistence. 

On \emph{simulated data}, we showed that persistence is more robust to zero-inflation than Moran's I, showing higher AUPRC and a better detection rate of truly variable features amongst the highest ranked genes. At a fixed significance cut-off, we further find that persistence-based methods lead to slightly lower sensitivity but higher specificity, suggesting that persistence is potentially more useful than Moran's I in applications that require more conservative estimates. We thus demonstrate in-silico that persistence curves offer stable approaches to quantify spatial dependence in feature values.
On \emph{real data}, we further observe that persistence is less dependent on the sum of feature values than Moran's I, indicating that it summarises spatial patterns in a more robust manner. 
Finally, we visualise genes with representative spatial patterns that are both spatially and highly variable and describe their qualitative biological relevance. Thus, we describe the practical benefits of using persistence-based approaches over Moran's I and showcase examples where persistence curves are superior to baseline methods.

Overall, we find that a non-parametric approach is less reliant on specific distributional assumptions about the spatial distribution of feature values, in contrast to parametric methods such as \texttt{SpatialDE}. This makes our approach preferable when such assumptions are not satisfied \citep{pesarin2010permutation}, which arguably remains a contested question for omics data in general. 
While our analysis focuses on illustrating examples on the use of persistent homology in the domain of spatial omics data analysis, our methodology can be generalised to other types of spatial data, such as spatial graphs with labelled vertices or edges. We thus propose a flexible method for detecting spatially variable features based on application-driven filtrations.

However, we want to also note the limitations of our study. The computational costs of non-parametric methods, especially the computation-heavy nature of permutation testing, remains a concern in practice \citep{pesarin2010permutation}.
Computing a persistence curve for one spatial feature itself is extremely fast, taking less than one second on a local CPU. 
However, the proposed spatial variability tests pose some scalability constraints. Permutation testing requires thousands of repetitions, resulting in a notable increase in computational costs \citep{li2023benchmarking}. While per feature this test might then take less than a minute, this scaling behaviour does not suffice when the goal is to study hundreds of thousands of genes simultaneously. Sequential permutation testing could be employed to reduce some of these costs \citep{hapfelmeier2023efficient}, but might not lead to sufficient computational improvements overall. There thus exists a great motivation for further work to develop alternative statistical tests using potentially parametric results, such as ones that could be derived from a universal null distribution for persistence diagrams \citep{bobrowski2023universal} to speed up the proposed testing procedures. 
%
%

Further, we have observed a high degree of disagreement between SVG detection methods across our experiments, which is a known practical issue for benchmarking these methods \citep{li2023benchmarking, chen2024evaluating, charitakis2023disparities}. Even across our simulation study, we find  no one spatial variability detection method that is preferable in \emph{all} scenarios. Indeed, further investigation could be beneficial to gain a more complete understanding of the advantages and disadvantages of each approach. 
Nevertheless, we conclude that persistence-based approaches offer a valid alternative to conventional spatial variability detection methods, in particular Moran's I, as they achieve high AUPRC values even in the presence of zero-inflation, give more specific results than Moran's I, and show lower correlation to the total sum of feature values. 

In terms of the way persistence is used throughout this study, we investigated zero-dimensional persistent homology because of its computational efficiency, as well as the utility of encoding local peaks for detecting spatial patterns. However, we note that one could follow an approach similar to \citet{hickok2022analysis} to extend the filtration by its exterior and compute one-dimensional persistent homology. This would permit deducing information about the geographical location of local minima via tracking the generators of one-dimensional homology classes, which could be combined with statistical methods to quantify exactly how spatial patterns vary from spatial randomness.


\section*{Acknowledgments}

The authors are grateful for the constructive and detailed feedback given by the reviewers. Their questions and feedback greatly helped in revising this manuscript. We thank the reviewers for believing in the merits of this work. 
We further want to thank Mairi McClean, Michaela Müller, Emily Simons, and Jera Toporiš for their help in proofreading the paper. 
K.L.\ is supported by the Helmholtz Association under the joint research school `Munich School for Data Science~(MUDS).'
B.R.\ was partially supported by the Bavarian state government with funds from the \emph{Hightech Agenda Bavaria}.
This work has received funding from the Swiss State
Secretariat for Education, Research, and Innovation~(SERI).
\newpage
\ifpreprint
\bibliographystyle{abbrvnat}
\fi
\bibliography{sn-article}

\begin{thebibliography}{47}
\providecommand{\natexlab}[1]{#1}
\providecommand{\url}[1]{\texttt{#1}}
\expandafter\ifx\csname urlstyle\endcsname\relax
  \providecommand{\doi}[1]{doi: #1}\else
  \providecommand{\doi}{doi: \begingroup \urlstyle{rm}\Url}\fi

\bibitem[Adhikari et~al.(2024)Adhikari, Yang, Wang, and Cui]{adhikari2024recent}
S.~D. Adhikari, J.~Yang, J.~Wang, and Y.~Cui.
\newblock Recent advances in spatially variable gene detection in spatial transcriptomics.
\newblock \emph{Computational and Structural Biotechnology Journal}, 2024.

\bibitem[Ahlmann-Eltze and Huber(2023)]{ahlmann2023comparison}
C.~Ahlmann-Eltze and W.~Huber.
\newblock Comparison of transformations for single-cell rna-seq data.
\newblock \emph{Nature Methods}, 20\penalty0 (5):\penalty0 665--672, 2023.

\bibitem[Aktas et~al.(2019)Aktas, Akbas, and Fatmaoui]{aktas2019persistence}
M.~E. Aktas, E.~Akbas, and A.~E. Fatmaoui.
\newblock Persistence homology of networks: methods and applications.
\newblock \emph{Applied Network Science}, 4\penalty0 (1):\penalty0 1--28, 2019.

\bibitem[Ali et~al.(2023)Ali, Asaad, Jimenez, Nanda, Paluzo-Hidalgo, and Soriano-Trigueros]{ali2023survey}
D.~Ali, A.~Asaad, M.-J. Jimenez, V.~Nanda, E.~Paluzo-Hidalgo, and M.~Soriano-Trigueros.
\newblock A survey of vectorization methods in topological data analysis.
\newblock \emph{IEEE Transactions on Pattern Analysis and Machine Intelligence}, 2023.

\bibitem[Am{\'e}zquita et~al.(2020)Am{\'e}zquita, Quigley, Ophelders, Munch, and Chitwood]{amezquita2020shape}
E.~J. Am{\'e}zquita, M.~Y. Quigley, T.~Ophelders, E.~Munch, and D.~H. Chitwood.
\newblock The shape of things to come: Topological data analysis and biology, from molecules to organisms.
\newblock \emph{Developmental Dynamics}, 249\penalty0 (7):\penalty0 816--833, 2020.

\bibitem[Andersson and Lundeberg(2021)]{andersson2021sepal}
A.~Andersson and J.~Lundeberg.
\newblock sepal: Identifying transcript profiles with spatial patterns by diffusion-based modeling.
\newblock \emph{Bioinformatics}, 37\penalty0 (17):\penalty0 2644--2650, 2021.

\bibitem[Benjamini and Hochberg(1995)]{benjamini1995controlling}
Y.~Benjamini and Y.~Hochberg.
\newblock Controlling the false discovery rate: a practical and powerful approach to multiple testing.
\newblock \emph{Journal of the Royal statistical society: series B (Methodological)}, 57\penalty0 (1):\penalty0 289--300, 1995.

\bibitem[Bobrowski and Skraba(2023)]{bobrowski2023universal}
O.~Bobrowski and P.~Skraba.
\newblock A universal null-distribution for topological data analysis.
\newblock \emph{Scientific Reports}, 13\penalty0 (1):\penalty0 12274, 2023.

\bibitem[Bubenik et~al.(2015)]{bubenik2015statistical}
P.~Bubenik et~al.
\newblock Statistical topological data analysis using persistence landscapes.
\newblock \emph{J. Mach. Learn. Res.}, 16\penalty0 (1):\penalty0 77--102, 2015.

\bibitem[Bucher et~al.(2020)Bucher, Martin, Jonietz, Raubal, and Westerholt]{bucher2020estimation}
D.~Bucher, H.~Martin, D.~Jonietz, M.~Raubal, and R.~Westerholt.
\newblock Estimation of moran’s i in the context of uncertain mobile sensor measurements.
\newblock In \emph{11th International Conference on Geographic Information Science (GIScience 2021)-Part I (2020)}. Schloss-Dagstuhl-Leibniz Zentrum f{\"u}r Informatik, 2020.

\bibitem[Byers et~al.(2023)Byers, Pritchard, Turner, and Weighill]{byers_detecting_2023}
S.~Byers, N.~Pritchard, J.~Turner, and T.~Weighill.
\newblock Detecting spatial dependence with persistent homology.
\newblock \emph{Nonlinear Theory and Its Applications, IEICE}, 14\penalty0 (2):\penalty0 106--125, 2023.

\bibitem[Charitakis et~al.(2023)Charitakis, Salim, Piers, Watt, Porrello, Elliott, and Ramialison]{charitakis2023disparities}
N.~Charitakis, A.~Salim, A.~T. Piers, K.~I. Watt, E.~R. Porrello, D.~A. Elliott, and M.~Ramialison.
\newblock Disparities in spatially variable gene calling highlight the need for benchmarking spatial transcriptomics methods.
\newblock \emph{Genome Biology}, 24\penalty0 (1):\penalty0 209, 2023.

\bibitem[Chen et~al.(2024{\natexlab{a}})Chen, Kim, and Yang]{chen2024evaluating}
C.~Chen, H.~J. Kim, and P.~Yang.
\newblock Evaluating spatially variable gene detection methods for spatial transcriptomics data.
\newblock \emph{Genome Biology}, 25\penalty0 (1):\penalty0 18, 2024{\natexlab{a}}.

\bibitem[Chen et~al.(2024{\natexlab{b}})Chen, Kim, and Yang]{chen_evaluating_2022}
C.~Chen, H.~J. Kim, and P.~Yang.
\newblock Evaluating spatially variable gene detection methods for spatial transcriptomics data.
\newblock \emph{Genome Biology}, 25\penalty0 (1):\penalty0 18, 2024{\natexlab{b}}.

\bibitem[Edelsbrunner et~al.(2008)Edelsbrunner, Harer, et~al.]{edelsbrunner2008persistent}
H.~Edelsbrunner, J.~Harer, et~al.
\newblock Persistent homology-a survey.
\newblock \emph{Contemporary mathematics}, 453\penalty0 (26):\penalty0 257--282, 2008.

\bibitem[Edsg{\"a}rd et~al.(2018)Edsg{\"a}rd, Johnsson, and Sandberg]{edsgard2018identification}
D.~Edsg{\"a}rd, P.~Johnsson, and R.~Sandberg.
\newblock Identification of spatial expression trends in single-cell gene expression data.
\newblock \emph{Nature methods}, 15\penalty0 (5):\penalty0 339--342, 2018.

\bibitem[Feng and Porter(2020)]{feng_spatial_2020}
M.~Feng and M.~A. Porter.
\newblock Spatial applications of topological data analysis: {Cities}, snowflakes, random structures, and spiders spinning under the influence.
\newblock \emph{Physical Review Research}, 2\penalty0 (3):\penalty0 033426, Sept. 2020.
\newblock Publisher: American Physical Society.

\bibitem[Geary(1954)]{geary1954contiguity}
R.~C. Geary.
\newblock The contiguity ratio and statistical mapping.
\newblock \emph{The incorporated statistician}, 5\penalty0 (3):\penalty0 115--146, 1954.

\bibitem[Hapfelmeier et~al.(2023)Hapfelmeier, Hornung, and Haller]{hapfelmeier2023efficient}
A.~Hapfelmeier, R.~Hornung, and B.~Haller.
\newblock Efficient permutation testing of variable importance measures by the example of random forests.
\newblock \emph{Computational Statistics \& Data Analysis}, 181:\penalty0 107689, 2023.

\bibitem[Heumos et~al.(2023)Heumos, Schaar, Lance, Litinetskaya, Drost, Zappia, L{\"u}cken, Strobl, Henao, Curion, et~al.]{heumos2023best}
L.~Heumos, A.~C. Schaar, C.~Lance, A.~Litinetskaya, F.~Drost, L.~Zappia, M.~D. L{\"u}cken, D.~C. Strobl, J.~Henao, F.~Curion, et~al.
\newblock Best practices for single-cell analysis across modalities.
\newblock \emph{Nature Reviews Genetics}, 24\penalty0 (8):\penalty0 550--572, 2023.

\bibitem[Hickok et~al.(2022)Hickok, Needell, and Porter]{hickok2022analysis}
A.~Hickok, D.~Needell, and M.~A. Porter.
\newblock Analysis of spatial and spatiotemporal anomalies using persistent homology: Case studies with covid-19 data.
\newblock \emph{SIAM Journal on Mathematics of Data Science}, 4\penalty0 (3):\penalty0 1116--1144, 2022.

\bibitem[Huber(2021)]{huber2021persistent}
S.~Huber.
\newblock Persistent homology in data science.
\newblock In \emph{Data Science--Analytics and Applications: Proceedings of the 3rd International Data Science Conference--iDSC2020}, pages 81--88. Springer, 2021.

\bibitem[Kuppe et~al.(2022)Kuppe, Ramirez~Flores, Li, Hayat, Levinson, Liao, Hannani, Tanevski, Wünnemann, Nagai, Halder, Schumacher, Menzel, Schäfer, Hoeft, Cheng, Ziegler, Zhang, Peisker, Kaesler, Saritas, Xu, Kassner, Gummert, Morshuis, Amrute, Veltrop, Boor, Klingel, Van~Laake, Vink, Hoogenboezem, Bindels, Schurgers, Sattler, Schapiro, Schneider, Lavine, Milting, Costa, Saez-Rodriguez, and Kramann]{kuppe_spatial_2022}
C.~Kuppe, R.~O. Ramirez~Flores, Z.~Li, S.~Hayat, R.~T. Levinson, X.~Liao, M.~T. Hannani, J.~Tanevski, F.~Wünnemann, J.~S. Nagai, M.~Halder, D.~Schumacher, S.~Menzel, G.~Schäfer, K.~Hoeft, M.~Cheng, S.~Ziegler, X.~Zhang, F.~Peisker, N.~Kaesler, T.~Saritas, Y.~Xu, A.~Kassner, J.~Gummert, M.~Morshuis, J.~Amrute, R.~J.~A. Veltrop, P.~Boor, K.~Klingel, L.~W. Van~Laake, A.~Vink, R.~M. Hoogenboezem, E.~M.~J. Bindels, L.~Schurgers, S.~Sattler, D.~Schapiro, R.~K. Schneider, K.~Lavine, H.~Milting, I.~G. Costa, J.~Saez-Rodriguez, and R.~Kramann.
\newblock Spatial multi-omic map of human myocardial infarction.
\newblock \emph{Nature}, 608\penalty0 (7924):\penalty0 766--777, Aug. 2022.
\newblock ISSN 1476-4687.

\bibitem[Li et~al.(2024)Li, Wang, and Lin]{li2024sgcast}
J.~Li, J.~Wang, and Z.~Lin.
\newblock Sgcast: symmetric graph convolutional auto-encoder for scalable and accurate study of spatial transcriptomics.
\newblock \emph{Briefings in Bioinformatics}, 25\penalty0 (1):\penalty0 bbad490, 2024.

\bibitem[Li et~al.(2023)Li, Patel, Song, Yan, Li, and Pinello]{li2023benchmarking}
Z.~Li, Z.~M. Patel, D.~Song, G.~Yan, J.~J. Li, and L.~Pinello.
\newblock Benchmarking computational methods to identify spatially variable genes and peaks.
\newblock \emph{Biorxiv}, pages 2023--12, 2023.

\bibitem[Moran(1950)]{moran1950notes}
P.~A. Moran.
\newblock Notes on continuous stochastic phenomena.
\newblock \emph{Biometrika}, 37\penalty0 (1/2):\penalty0 17--23, 1950.

\bibitem[Moses and Pachter(2022)]{moses2022museum}
L.~Moses and L.~Pachter.
\newblock Museum of spatial transcriptomics.
\newblock \emph{Nature methods}, 19\penalty0 (5):\penalty0 534--546, 2022.

\bibitem[Palla et~al.(2022{\natexlab{a}})Palla, Fischer, Regev, and Theis]{palla_spatial_2022}
G.~Palla, D.~S. Fischer, A.~Regev, and F.~J. Theis.
\newblock Spatial components of molecular tissue biology.
\newblock \emph{Nature Biotechnology}, 40\penalty0 (3):\penalty0 308--318, Mar. 2022{\natexlab{a}}.
\newblock ISSN 1546-1696.

\bibitem[Palla et~al.(2022{\natexlab{b}})Palla, Spitzer, Klein, Fischer, Schaar, Kuemmerle, Rybakov, Ibarra, Holmberg, Virshup, et~al.]{palla2022squidpy}
G.~Palla, H.~Spitzer, M.~Klein, D.~Fischer, A.~C. Schaar, L.~B. Kuemmerle, S.~Rybakov, I.~L. Ibarra, O.~Holmberg, I.~Virshup, et~al.
\newblock Squidpy: a scalable framework for spatial omics analysis.
\newblock \emph{Nature methods}, 19\penalty0 (2):\penalty0 171--178, 2022{\natexlab{b}}.

\bibitem[Pereira and de~Mello(2015)]{pereira2015persistent}
C.~M. Pereira and R.~F. de~Mello.
\newblock Persistent homology for time series and spatial data clustering.
\newblock \emph{Expert Systems with Applications}, 42\penalty0 (15-16):\penalty0 6026--6038, 2015.

\bibitem[Pesarin and Salmaso(2010)]{pesarin2010permutation}
F.~Pesarin and L.~Salmaso.
\newblock The permutation testing approach: a review.
\newblock \emph{Statistica}, 70\penalty0 (4):\penalty0 481--509, 2010.

\bibitem[Phipson and Smyth(2010)]{phipson2010permutation}
B.~Phipson and G.~K. Smyth.
\newblock Permutation p-values should never be zero: calculating exact p-values when permutations are randomly drawn.
\newblock \emph{Statistical applications in genetics and molecular biology}, 9\penalty0 (1), 2010.

\bibitem[Rabad{\'a}n and Blumberg(2019)]{rabadan2019topological}
R.~Rabad{\'a}n and A.~J. Blumberg.
\newblock \emph{Topological data analysis for genomics and evolution: topology in biology}.
\newblock Cambridge University Press, Cambridge, November 2019.
\newblock ISBN 9781316671665.

\bibitem[Reel et~al.(2021)Reel, Reel, Pearson, Trucco, and Jefferson]{reel2021using}
P.~S. Reel, S.~Reel, E.~Pearson, E.~Trucco, and E.~Jefferson.
\newblock Using machine learning approaches for multi-omics data analysis: A review.
\newblock \emph{Biotechnology advances}, 49:\penalty0 107739, 2021.

\bibitem[Rieck and Leitte(2016)]{Rieck16a}
B.~Rieck and H.~Leitte.
\newblock Exploring and comparing clusterings of multivariate data sets using persistent homology.
\newblock \emph{Computer Graphics Forum}, 35\penalty0 (3):\penalty0 81--90, 2016.

\bibitem[Rieck et~al.(2017)Rieck, Fugacci, Lukasczyk, and Leitte]{rieck2017clique}
B.~Rieck, U.~Fugacci, J.~Lukasczyk, and H.~Leitte.
\newblock Clique community persistence: A topological visual analysis approach for complex networks.
\newblock \emph{IEEE transactions on visualization and computer graphics}, 24\penalty0 (1):\penalty0 822--831, 2017.

\bibitem[Rieck et~al.(2020)Rieck, Sadlo, and Leitte]{rieck2020topological}
B.~Rieck, F.~Sadlo, and H.~Leitte.
\newblock Topological machine learning with persistence indicator functions.
\newblock In \emph{Topological Methods in Data Analysis and Visualization V: Theory, Algorithms, and Applications 7}, pages 87--101, Heidelberg, 2020. Springer.

\bibitem[Robinson and Turner(2017)]{robinson_hypothesis_2017}
A.~Robinson and K.~Turner.
\newblock Hypothesis testing for topological data analysis.
\newblock \emph{Journal of Applied and Computational Topology}, 1\penalty0 (2):\penalty0 241--261, Dec. 2017.
\newblock ISSN 2367-1734.

\bibitem[St{\aa}hl et~al.(2016)St{\aa}hl, Salm{\'e}n, Vickovic, Lundmark, Navarro, Magnusson, Giacomello, Asp, Westholm, Huss, et~al.]{staahl2016visualization}
P.~L. St{\aa}hl, F.~Salm{\'e}n, S.~Vickovic, A.~Lundmark, J.~F. Navarro, J.~Magnusson, S.~Giacomello, M.~Asp, J.~O. Westholm, M.~Huss, et~al.
\newblock Visualization and analysis of gene expression in tissue sections by spatial transcriptomics.
\newblock \emph{Science}, 353\penalty0 (6294):\penalty0 78--82, 2016.

\bibitem[Storey and Tibshirani(2003)]{storey2003statistical}
J.~D. Storey and R.~Tibshirani.
\newblock Statistical significance for genomewide studies.
\newblock \emph{Proceedings of the National Academy of Sciences}, 100\penalty0 (16):\penalty0 9440--9445, 2003.

\bibitem[Svensson et~al.(2018)Svensson, Teichmann, and Stegle]{svensson_spatialde_2018}
V.~Svensson, S.~A. Teichmann, and O.~Stegle.
\newblock {SpatialDE}: identification of spatially variable genes.
\newblock \emph{Nature Methods}, 15\penalty0 (5):\penalty0 343--346, May 2018.
\newblock ISSN 1548-7105.

\bibitem[Tian et~al.(2023)Tian, Chen, and Macosko]{tian_expanding_2022}
L.~Tian, F.~Chen, and E.~Z. Macosko.
\newblock The expanding vistas of spatial transcriptomics.
\newblock \emph{Nature Biotechnology}, 41\penalty0 (6):\penalty0 773--782, 2023.

\bibitem[Tiefelsdorf and Boots(1997)]{tiefelsdorf1997note}
M.~Tiefelsdorf and B.~Boots.
\newblock A note on the extremities of local moran's iis and their impact on global moran's i.
\newblock \emph{Geographical Analysis}, 29\penalty0 (3):\penalty0 248--257, 1997.

\bibitem[Van~Rooij et~al.(2008)Van~Rooij, Sutherland, Thatcher, DiMaio, Naseem, Marshall, Hill, and Olson]{van2008dysregulation}
E.~Van~Rooij, L.~B. Sutherland, J.~E. Thatcher, J.~M. DiMaio, R.~H. Naseem, W.~S. Marshall, J.~A. Hill, and E.~N. Olson.
\newblock Dysregulation of micrornas after myocardial infarction reveals a role of mir-29 in cardiac fibrosis.
\newblock \emph{Proceedings of the National Academy of Sciences}, 105\penalty0 (35):\penalty0 13027--13032, 2008.

\bibitem[Vandereyken et~al.(2023)Vandereyken, Sifrim, Thienpont, and Voet]{vandereyken2023methods}
K.~Vandereyken, A.~Sifrim, B.~Thienpont, and T.~Voet.
\newblock Methods and applications for single-cell and spatial multi-omics.
\newblock \emph{Nature Reviews Genetics}, 24\penalty0 (8):\penalty0 494--515, 2023.

\bibitem[Wagner et~al.(2011)Wagner, Chen, and Vu{\c{c}}ini]{wagner2011efficient}
H.~Wagner, C.~Chen, and E.~Vu{\c{c}}ini.
\newblock Efficient computation of persistent homology for cubical data.
\newblock In \emph{Topological methods in data analysis and visualization II: theory, algorithms, and applications}, pages 91--106. Springer, Heidelberg, 2011.

\bibitem[Zhu et~al.(2023)Zhu, Shang, and Zhou]{zhu2023srtsim}
J.~Zhu, L.~Shang, and X.~Zhou.
\newblock Srtsim: spatial pattern preserving simulations for spatially resolved transcriptomics.
\newblock \emph{Genome biology}, 24\penalty0 (1):\penalty0 39, 2023.

\end{thebibliography}

\end{document}